\documentclass[journal]{IEEEtran}

\IEEEoverridecommandlockouts
\usepackage{cite}
\usepackage{amsmath,amssymb,amsfonts,bm}
\usepackage{algorithmic}
\usepackage{graphicx}
\usepackage{textcomp}
\usepackage{caption}
\usepackage{subcaption}
\usepackage{xcolor}
\usepackage{tabularx}
\usepackage{booktabs}
\usepackage{multirow}
\usepackage{rotating}
\ifCLASSINFOpdf
\else
\fi

\begin{document}

\title{Task-Oriented Mulsemedia Communication using Unified Perceiver and Conformal Prediction in 6G {Wireless} Systems}

\author{{Hongzhi~Guo},~\IEEEmembership{Senior Member,~IEEE,}
        and~{Ian~F.~Akyildiz},~\IEEEmembership{Life~Fellow,~IEEE}
\thanks{H. Guo is with the University of Nebraska-Lincoln, Lincoln, NE, 68588, USA. E-mail: hguo10@unl.edu. I. F. Akyildiz is with Truva Inc., Alpharetta, GA 30022, USA. E-mail: ian@truvainc.com and with Odine Research Labs, Istanbul, Turkiye, E-mail: ian.akyildiz@odine.com }
\thanks{This work was supported in part by the Layman Award (Layman Seed Program) from the University of Nebraska Foundation.}
\thanks{}}


\maketitle

\begin{abstract}

The growing prominence of eXtended Reality (XR), holographic-type communications, and metaverse demands truly immersive user experiences by using many sensory modalities, including sight, hearing, touch, smell, taste, etc. Additionally, the widespread deployment of sensors in areas such as agriculture, manufacturing, and smart homes is generating diverse sensory data. A new media format known as multisensory media (mulsemedia) has emerged, which incorporates many sensory modalities beyond the traditional visual and auditory media. 6G wireless systems are envisioned to support the Internet of Senses, making it crucial to explore effective data fusion and communication strategies for mulsemedia. In this paper, we introduce a task-oriented multi-task mulsemedia communication system named MuSeCo, which is developed using unified Perceiver models and Conformal Prediction. This unified model can accept any sensory input and efficiently extract latent semantic features, making it adaptable for deployment across various Artificial Intelligence of Things (AIoT) devices. Conformal Prediction is employed for modality selection and combination, enhancing task accuracy while minimizing data communication overhead. The model is trained using six sensory modalities across four classification tasks. Simulations and experiments demonstrate that it can effectively fuse sensory modalities, significantly reduce end-to-end communication latency and energy consumption, and maintain high accuracy in communication-constrained systems.

\end{abstract}


\IEEEpeerreviewmaketitle

\section{Introduction}

Human perception is a sophisticated process that integrates multiple sensory modalities, including the five fundamental senses: vision, audition, tactile sensation, olfaction, and gustation. Recent neuroscientific discoveries have revealed the existence of a multisensory correlation detector in the human brain, which enables optimal selection and integration of sensory inputs to enhance perception \cite{pesnot2022multisensory, parise2016correlation}. This ability is not exclusive to humans; even simple organisms exhibit comparable sensory integration mechanisms. Historically, the simultaneous use of multiple sensory modalities for a single task has been technically challenging. Prior to the advent of 5G wireless systems, the transmission of novel sensory modalities, such as tactile, olfactory, and gustatory signals, was hindered by limitations in communication and sensing technologies. However, the emergence of 5G has significantly reduced end-to-end latency and increased data rates, paving the way for the transmission of advanced sensory modalities, including haptic signals via the tactile Internet \cite{antonakoglou2018toward}. With the progression toward 6G wireless systems, it is anticipated that even more sensory modalities will be integrated, advancing the vision of the ``Internet of Senses'' \cite{joda2022internet}. This evolution is poised to revolutionize eXtended Reality (XR), encompassing Augmented Reality (AR), Mixed Reality (MR), and Virtual Reality (VR), as well as Holographic-Type Communications (HTC), which leverage spatial computing to create hyperrealistic environments and objects \cite{akyildiz2022wireless, akyildiz2022holographic}. Furthermore, the development of digital twins for humans and virtual environments heavily depends on the integration of multiple sensory modalities. Head-Mounted Displays (HMDs) for XR and HTC are equipped with advanced sensors and actuators, which are critical for capturing and rendering diverse sensory data. As XR and HTC become integral components of metaverse applications, there is a growing demand for multisensory media (mulsemedia) that seamlessly integrates a wide range of sensory modalities \cite{covaci2018multimedia, da2022ongoing, akyildiz2023mulsemedia}.

Mulsemedia represents a significant advancement over traditional multimedia by incorporating multiple sensory modalities beyond the visual and auditory domains. Unlike multimodal datasets, which are static and pre-collected, mulsemedia is a dynamic framework focused on real-time streaming and interactive applications \cite{akyildiz2023mulsemedia}. {While existing multimodal data processing techniques can support mulsemedia systems, mulsemedia communication can be categorized into two distinct objectives. The first focuses on the precise reconstruction of mulsemedia at remote destinations, prioritizing reconstruction fidelity \cite{covaci2018multimedia}. The second emphasizes the application of mulsemedia for specific tasks, such as prediction, classification, and identification, where semantic fidelity is paramount. This paper addresses the latter category, concentrating on the development of task-oriented mulsemedia communication systems.}

\begin{figure}[t]
\centering
    \includegraphics[width=0.35\textwidth]{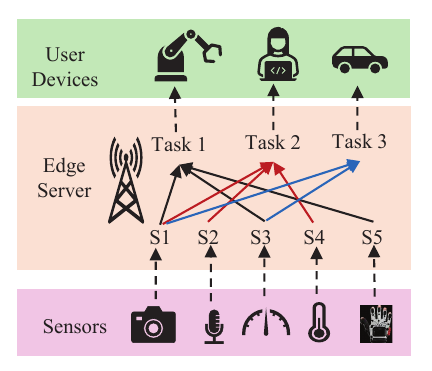}
    \vspace{-5pt}
    \caption{Illustration of mulsemedia communication and multi-task machine learning in edge networks. The sensors can capture senses including but not limited to sight, hearing, touch, smell, taste, temperature, moisture, and motion. }
    \label{fig:sys1}
    \vspace{-5pt}
\end{figure}

The proposed approach leverages an edge computing architecture, as illustrated in Fig.~\ref{fig:sys1}, where devices equipped with Artificial Intelligence (AI) modules and sensors transmit unimodal data to edge servers, such as gateways and base stations. These servers aggregate and process the unimodal data to generate mulsemedia, which is then transmitted to user devices for task execution. While existing Internet of Things (IoT) systems provide the foundation for this architecture, three significant research challenges persist, fundamentally limiting system performance in terms of task accuracy, end-to-end latency, and energy consumption. Addressing these challenges is essential to unlock the full potential of mulsemedia in enabling next-generation applications.

First, the inherent heterogeneity in data dimensions and quality across various sensory modalities presents a fundamental challenge in data encoding, communication, and fusion. For instance, visual data, such as videos and images, often dominate in terms of volume, while data from sources like pressure sensors and motion tracking systems are comparatively sparse. In communication systems that rely on data packets, prioritizing smaller sensory modalities becomes critical. A failure to do so could result in the complete loss of an entire sensory modality due to the loss of a single data packet. Additionally, sensory data captured at different times necessitates advanced synchronization techniques. The presence of low-quality or misaligned sensory data can significantly degrade the performance of data fusion processes and, consequently, the outcomes of downstream tasks.

\begin{figure}[t]
\centering
    \includegraphics[width=0.35\textwidth]{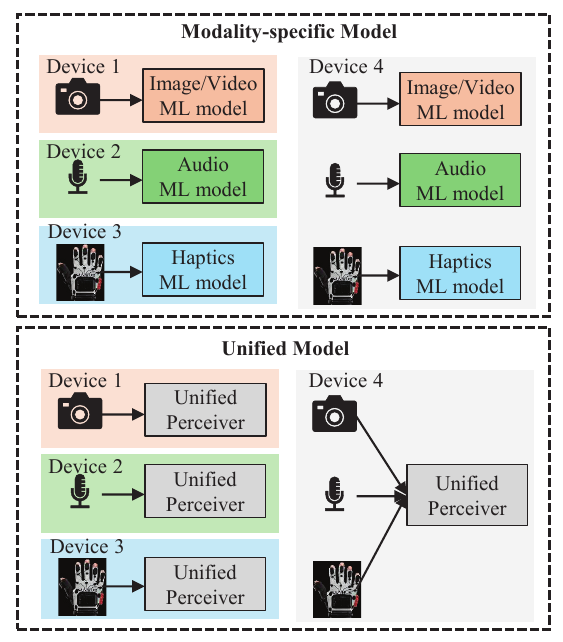}
    \vspace{-5pt}
    \caption{{Illustration of modality-specific machine learning model and unified machine learning model. A single device can be equipped with multiple modality-specific models, each tailored to process different sensory modalities, such as the five human basic senses: sight, hearing, touch, smell, and taste. Differently, a unified model can be designed to handle multiple sensory modalities and perform various tasks, including those related to the same five senses.}}
    \label{fig:specified_model}
    \vspace{-5pt}
\end{figure}

Second, IoT devices exhibit considerable heterogeneity in their computing, storage, and communication capabilities, further complicating the processing of multimodal data. Traditional approaches to multimodal data processing employ modality-specific machine learning (ML) models. For example, images and audio are often processed using separate, independent models to extract latent features, which are then fused to generate task-specific outputs. However, in the context of mulsemedia systems, which involve multiple sensory modalities, this approach places a significant burden on IoT devices. A simple IoT device may be required to support numerous modality-specific models, potentially exceeding its computational and storage capacities, as illustrated in Fig.~\ref{fig:specified_model}. Moreover, feature extraction models are often task-specific, meaning that employing a single sensory modality for multiple tasks necessitates the use of a variety of models. Developing a unified solution capable of processing arbitrary sensory modalities for multiple tasks remains a formidable challenge.

Last, a sensory modality may lack information relevant to a specific task, or such information may be distorted by noise or interference. In traditional multimodal ML systems, data is preprocessed and carefully selected to ensure its relevance to the task at hand; for example, synchronizing a text file with an audio file. However, in real-time streaming scenarios, an audio file may contain background noises that obscure the content, and in severe cases, render the associated data irrelevantly. Therefore, it is essential to assess the task-related information present in each sensory modality and intelligently combine these modalities to optimize task performance. This requires a dynamic system capable of evaluating and integrating sensory data in real time, effectively addressing these challenges and ensuring robust performance.

To address the aforementioned challenges, our ambitious objective is to develop a unified ML model capable of encoding diverse sensory modalities into a compact, homogeneous latent representation suitable for efficient transmission. This model will also incorporate adaptive mechanisms to selectively prioritize essential sensory modalities, thereby minimizing end-to-end communication latency, reducing energy consumption, and maximizing task accuracy.

First, we propose a distributed and unified mulsemedia communication and data fusion architecture, named MuSeCo (MulSemedia Communication), employing the Perceiver model \cite{jaegle2021perceiver} and the HighMMT multimodal transformer \cite{liang2022high}. This architecture allows for the processing of sensory modalities of arbitrary data dimensions through the unified Perceiver model, which transforms these inputs into homogeneous latent data of significantly reduced size compared to the raw data. Notably, the Perceiver model, in contrast to the traditional Transformer \cite{vaswani2017attention}, can accommodate inputs of any dimension and often requires fewer parameters due to its reuse of attention layers. MuSeCo can be seamlessly integrated into Artificial Intelligence of Things (AIoT) devices equipped with various sensors; for instance, both a camera and a microphone can operate using the same trained MuSeCo model without necessitating any modality-specific modifications. As illustrated in Fig.~\ref{fig:specified_model}, each AIoT device is equipped with a unified Perceiver module, eliminating the need for multiple modality-specific ML models in devices with multiple sensors.

Second, we significantly enhance the MuSeCo framework by integrating a sensory modality selection and combination mechanism that leverages Conformal Prediction \cite{angelopoulos2023conformal}. Drawing inspiration from the human brain's multisensory correlation detection capabilities, as demonstrated in \cite{pesnot2022multisensory}, our approach differs fundamentally from the HighMMT model, which does not have an explicit evaluation of the role and contribution of individual sensory modalities. In contrast, MuSeCo employs Conformal Prediction to quantify and evaluate the correlation among different sensory modalities, dynamically adjusting their individual weights in the final output. Modalities that exhibit higher confidence levels and demonstrate greater relevance to the task at hand are assigned proportionally larger weights, ensuring a more robust and task-oriented integration of sensory data. To further optimize the fusion of multisensory information, we propose two innovative combination approaches: the equal-weight combination (EWC) and the set-size-scaled combination (SSSC). A schematic diagram of the proposed MuSeCo architecture is illustrated in Fig.~\ref{fig:sys}.

Additionally, we present an adaptive extension of our framework, MuSeCo-Adaptive, which enables the continuous transmission of Conformal Prediction scores and latent semantic data. This adaptive innovation is specifically designed to achieve a balance among three critical factors: minimal end-to-end latency, reduced energy consumption, and the maintenance of high task accuracy. These objectives are particularly crucial in systems that operate under strict data communication constraints. To validate the effectiveness of both MuSeCo and MuSeCo-Adaptive, we evaluate their performance using four diverse multimodal datasets that collectively encompass six distinct sensory modalities: image, audio, text, force, proprioception, and control signals. {The unified architecture of our model demonstrates exceptional versatility by successfully performing four distinct tasks without requiring task-specific training, thereby showcasing its adaptability and generalization capabilities.}

The remainder of this paper is organized as follows: Section II provides a comprehensive review of related work in the field. Section III details the proposed MuSeCo architecture, including the development of the Perceiver model and the design of the distributed communication system. Section IV delves into the methodologies for adaptive modality selection and combination, while also introducing the data communication protocols employed. Section V presents a thorough experimental evaluation of the system's performance through rigorous simulations. Finally, Section VI concludes the paper with a summary of our contributions and insights for future research.


\section{Related Work}
Two research domains relevant to this paper are semantic communication and multimodal machine learning. In semantic communication, rather than transmitting conventional bit streams, transmitters encode and send semantic data to enhance reliability and minimize communication overhead \cite{shi2021semantic}. {This approach serves two primary purposes. First, it facilitates data reconstruction where the transmitter encodes semantic information which enables the receiver to reconstruct the same modality with high-precision semantic details. This approach has been successfully implemented in various media including text \cite{yan2022resource,xie2021deep}, images and videos \cite{jiang2022wireless}, and audio \cite{weng2021semantic}. Second, semantic communication is employed in task-oriented contexts such as image retrieval, machine translation, and visual question answering \cite{xie2022task,xie2021task}. Task-oriented semantic communication significantly reduces communication overhead by requiring only the transmission of task-relevant information, thereby reducing the volume of encoded data.}

Previous research has also delved into multimodal semantic communication. For instance, in \cite{xie2021task}, image and text modalities are integrated for the visual question answering task. The work in \cite{wang2024distributed} addresses the audio-visual parsing task, while \cite{zhang2024unified} considers modalities including text, image, and speech. These studies typically employ modality-specific encoders, i.e., each sensory modality is processed using an encoder specifically designed for that type which is widely used in multimodal data fusion \cite{zhang2023multimodal}. This paper distinguishes itself from the existing literature by utilizing a unified encoder based on the Perceiver model, which eliminates the need for modality-specific encoders. This unified encoder is capable of handling arbitrary types of modalities, regardless of their input dimensions. Thus, it is not confined to conventional modalities such as text, image, or speech, offering a more versatile and comprehensive approach to multimodal semantic communication.

The design of the proposed MuSeCo architecture is inspired by the advancements in general intelligence for multimodal agents \cite{durante2024agent}. {There is a trend in the development of AI agents that possess general intelligence capabilities which enable them to process multimodal data across multiple tasks. For instance, the study in \cite{hu2021unit} reports on a unified Transformer model trained to concurrently learn 7 tasks over 8 datasets, primarily utilizing visual and language modalities. Similarly, the works in \cite{liang2022high,liang2023multizoo} feature unified Transformer models that handle 10 modalities for 15 tasks. This shows that the unified model architecture using Transformer or Perceiver can generalize well to multiple tasks.} While these models achieve state-of-the-art (SOTA) performance with fewer parameters, their architectures are centralized, which limits their applicability in distributed data communication environments. In contrast, the MuSeCo architecture reorganizes and distributes the attention modules in Perceiver \cite{jaegle2021perceiver} and HighMMT \cite{liang2022high} to AIoT devices and edge servers. This decentralized approach facilitates more flexible and scalable applications in real-world environments. Additionally, unlike the Perceiver and HighMMT models that rely on static, pre-collected data, MuSeCo incorporates a modality selection and combination module that adaptively processes available sensory modalities. This feature enables dynamic handling of data inputs, enhancing the system's efficacy in diverse operational contexts.

Existing work \cite{cai2023acf} also studied adaptive modality fusion with focus on compression. Differently, in this paper the modality selection and combination process within the MuSeCo architecture utilizes Conformal Prediction, which offers prediction certainty scores and sets, rather than merely providing a straightforward prediction result \cite{shafer2008tutorial, angelopoulos2021uncertainty}. While ML models are capable of generating outputs for classification tasks, these outputs typically lack information regarding the certainty of the predictions. Conformal Prediction addresses this limitation by generating a classification set, denoted as ${\bm u}$, along with a confidence level $1-\alpha$. This ensures that the coverage of Conformal Prediction will include the true class within ${\bm u}$ with a probability of $1-\alpha$. Both theoretical analyses and practical implementations have demonstrated the use of Conformal Prediction in enhancing ML systems \cite{barber2023conformal, angelopoulos2021uncertainty, angelopoulos2023conformal}. In this paper, we leverage the Softmax outputs from the Perceiver model as the basis for generating conformal prediction results. These results are then utilized to select sensory modalities and determine optimal modality combination coefficients. This method enriches the predictive reliability and efficacy of the system by integrating a quantitative measure of confidence into the decision-making process, thereby enhancing the overall performance of the multimodal integration.
  
\section{System Architecture and Modeling}
\begin{figure*}[t]
\centering
    \includegraphics[width=0.75\textwidth]{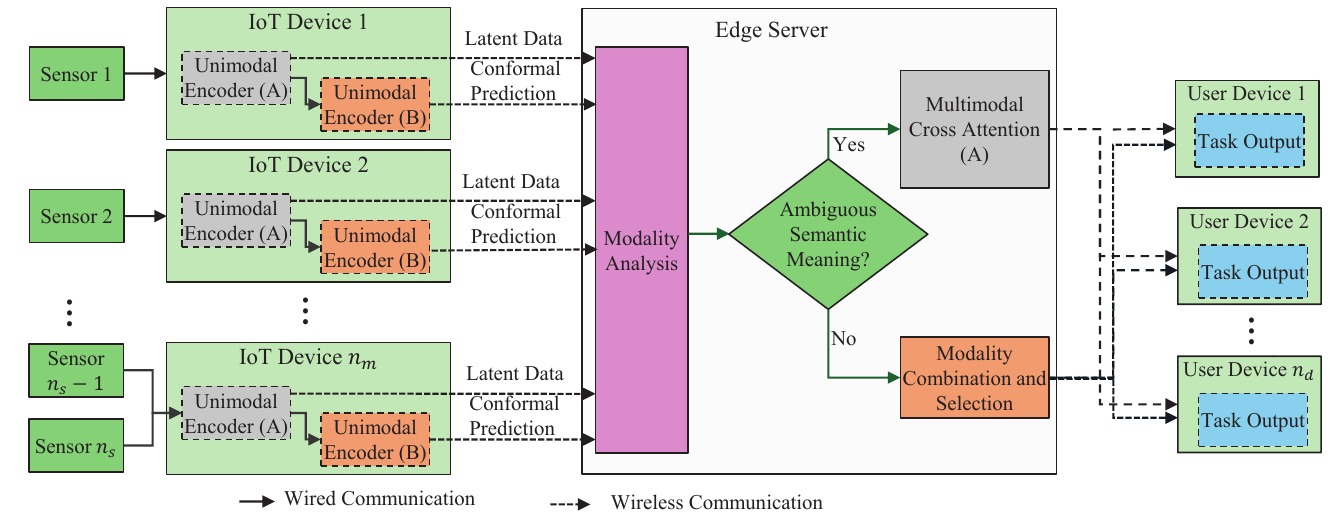}
    \vspace{-5pt}
    \caption{MuSeCo system architecture using unified Perceiver encoder and Conformal Prediction for Artificial Intelligence of Things.}
    \label{fig:sys}
    \vspace{-5pt}
\end{figure*}

In this section, we present the architecture of the proposed MuSeCo system, specifically designed for seamless deployment across distributed AIoT environments. To complement this architecture, we also propose an analytical model that provides a comprehensive description of the system's operational mechanisms. As illustrated in Fig.~\ref{fig:sys}, the MuSeCo framework is composed of $n_m$ AIoT devices, such as Raspberry Pis, each equipped with one or more sensors to capture relevant data from their surroundings. Additionally, the system integrates $n_d$ user devices, each tailored to execute distinct applications, enabling versatile functionality. At the edge of the network, a gateway or base station is deployed to host an edge server. This server serves as the central processing unit, responsible for several critical tasks: it collects sensory data from the AIoT devices, performs advanced mulsemedia modality analysis, and generates accurate classification outputs. The system's design emphasizes scalability, efficiency, and robustness, making it well-suited for diverse AIoT applications.

\begin{table}
\centering
    \caption{Mathematical Notation.}
\label{tab:notation}
\begin{tabularx}{\linewidth}{l  X}
    \toprule
    Symbol  &  Description  \\
    \midrule
   ${\bm M}$ & Sensory modality data\\
   $n_m$& Number of AIoT devices \\
   $T_b$ & Computation load of Perceiver module\\
   $n_{t,i}$ &Number of sensory modalities for Task $i$\\
   ${\bm p}$ & Position embedding  \\
   ${\bm e}$ & Modality embedding \\
   ${\bm L}$ & Latent semantic data \\
   ${\bm c}$ & Softmax score \\
   ${\bm u}$ & Conformal prediction set \\
   $c_s$ & Server computing capacity \\
   $c_{iot}$ & IoT device computing capacity \\
   $r_t$ & Communication data rate \\
   $P_t$ & Communication transmission power \\
   $\Gamma_s$ & Computing energy efficiency of the server \\
   $\Gamma_{iot}$ & Computing energy efficiency of the IoT device\\
   $p_h$ & Percentage of data samples only use Conformal Prediction in MuSeCo-Adaptive\\
   $|{\bm M}|$ & The overall number of bits used to represent all data in ${\bm M}$ in binary. \\
    \bottomrule 
\end{tabularx}
\end{table}

We consider a scenario with $n_t$ tasks, where the $i$-th task utilizes a distinct set of sensory modalities denoted by ${\mathcal T}_i = \{{\bm M}_{i,1}, {\bm M}_{i,2}, \ldots \}$, with ${\bm M}_{i,j}$ representing the $j$-th sensory modality data for the $i$-th task. Henceforth, unless otherwise noted, we omit the task subscript $i$, and the discussions are applicable to any task.


\subsection{Unimodal Encoder}
The architecture of MuSeCo is shown in Fig.~\ref{fig:sys} where MuSeCo distributes the unimodal feature encoding to AIoT devices and leaves the multimodal fusion in the edge server. This can reduce the communication overhead. In the following, we introduce the unimodal encoder in MuSeCo. 

A key challenge in feature encoding is that different modalities have drastically different data dimension and format. For example, an image and a force signal both can be represented in matrix format, but their dimensions are different. For example, as shown in Table~\ref{tab:modality}, an image in Task 2 is 20$\times$371 and a force signal in Task 4 is 16$\times$7. Moreover, images are organized as spatial pixels and force signals are time-series data. Due to the lack of unified tools, existing works use modality-specific ML models to extract features and fuse them in later stages \cite{zhang2024unified}. 

Transformer is widely used in image, text, and audio machine learning \cite{vaswani2017attention}. It is also adopted in multi-task multimodal semantic communication \cite{zhang2024unified}. However, Transformer models cannot effectively reduce input data dimension and data communication overhead. Thus, modality-specific Transformer and dimension reduction models must be developed to process individual modality. This is mainly due to the query-key-value (QKV) attention mechanism that is used in Transformer. 

Let ${\bm M}\in \mathbb{R}^{l_m \times d_m}$ denote the input sensory modality and $f_{sa}\left({\bm M}\right) \in \mathbb{R}^{l_m \times d_a}$ represent the self-attention operation on ${\bm M}$, where $d_a = d_m/h_m$ and $h_m$ is the number of self-attention module. In Transformer, $ f_{sa}(\cdot)$ is written as:
\begin{align}
    f_{sa}\left({\bm M}\right)=\operatorname{softmax}\left(\frac{{\bm Q}{\bm  K}^{T}}{\sqrt{d_a}}\right) {\bm V},
\end{align}
where ${\bm Q} \in \mathbb{R}^{l_m \times d_a}$, ${\bm K} \in \mathbb{R}^{l_m \times d_a}$, and ${\bm V} \in \mathbb{R}^{l_m \times d_a}$ , and $\operatorname{softmax}(\cdot)$ is the Softmax function. The QKV matrices are generated using linear projections:
$$
{\bm Q}={\bm M} {\bm W}_Q, \quad {\bm K}={\bm M} {\bm W}_K, \quad {\bm V}={\bm M} {\bm W}_V,
$$
where $\boldsymbol{W}_Q \in \mathbb{R}^{d_m \times d_a}, \boldsymbol{W}_K \in \mathbb{R}^{d_m \times d_a}, \boldsymbol{W}_V \in \mathbb{R}^{d_m \times d_a}$. The multi-head self-attention module can be modeled as:
\begin{align}
    f_{msa}({\bm M})={\bm M}+[f_{sa}({\bm M}),\cdots,f_{sa}({\bm M})]{\bm W},
\end{align}
where there are $h_m$ $f_{sa}({\bm M})$ in the bracket and ${\bm W}\in \mathbb{R}^{d_m \times d_m}$. The Transformer encoder generates output:
\begin{align}
    {\bm L}_{out} = f_{msa}({\bm M}) + f_{ffn}(f_{msa}({\bm M})),
\end{align}
where $f_{ffn}(\cdot)$ is the feedforward network. Note that, the output ${\bm L}_{out}$ and input ${\bm M}$ have the same dimension. As a result, if the input has a large size, the Transformer encoder cannot generate latent data with reduced dimension. For multimodal sensory data with different input dimensions, modality-specific Transformer model must be designed, such as the work in \cite{zhang2024unified}. Additional steps are needed to combine different sensory modalities after Transformer encoding.

\begin{figure}[t]
\centering
    \includegraphics[width=0.5\textwidth]{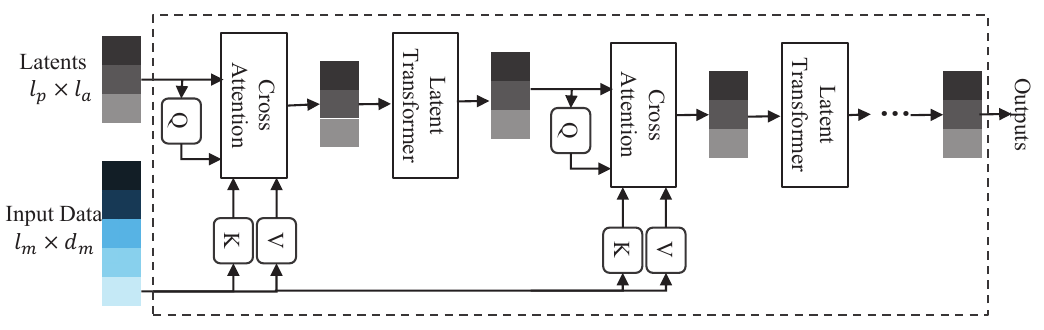}
    \vspace{-0pt}
    \caption{Illustration of the Perceiver architecture.}
    \label{fig:perceiver}
    \vspace{-5pt}
\end{figure}

Perceiver is developed by modifying Transformer \cite{jaegle2021perceiver}. In Transformer, the complexity of the self attention is ${\mathcal O}(l_m^2)$ which depends on the input size. To address this issue, the Perceiver uses asymmetric QKV attention module. In particular, ${\bm Q}$ becomes a projection of learned latent data with dimension $l_p$ and $K$ and $V$ are projections of the input data. Here, $l_p \ll l_m$ and complexity is reduced to ${\mathcal O}(l_p\cdot l_m)$. As shown in Fig.~\ref{fig:perceiver}, the output of the cross attention layer has the same dimension as the latent matrix and the complexity is reduced to ${\mathcal O}(l_p^2)$ in the latent space. The Perceiver uses a Fourier feature for position encoding and the small latent space allows deep learning with a large number of layers. Also, Perceiver reuses parameters in different layers which significantly reduces overall parameter number in the system. 

{In order to accept all input modalities, the input ${\bm M}$ is concatenated with one-hot modality embedding ${\bm e}$, position embedding ${\bm p}$ and zero padding to make all input modalities the same size, i.e., 
\begin{align}
    {\tilde {\bm M}} = [{\bm M}; {\bm e}; {\bm p}; {\bm 0}].
\end{align}
Therefore, based on all tasks and associated sensory modalities, we need to select the sensory modality with the largest size and use zero padding for other smaller modalities.} 

The Perceiver-based unimodal encoder can be written as:
\begin{align}
    {\bm L}_{u} = f_{uni} ({\tilde {\bm M}}, {\bm L}_0; \theta_A),  
\end{align}
where ${\bm L}_u \in {\mathbb R}^{l_p \times l_a}$ which is the output latent data, ${\bm L}_0\in {\mathbb R}^{l_p \times l_a}$ is the trained input latent data, and $\theta_A$ is the unimodal encoder parameters. Note that, ${\bm L}_u$ and ${\bm L}_0$ have the same dimension. The AIoT device can directly send ${\bm L}_u$ to the edge server for further multimodal fusion. Note that, compare to the raw sensory data, such as image and audio, ${\bm L}_u$ can be much smaller due to the reduced dimension of the latent space. For example, the latent dimension in Section~\ref{sec:eval} is 20$\times$64 which is much smaller than the image dimension in Task 2, i.e., $112\times 112$ as shown in Table~\ref{tab:modality}. This could reduce the data communication overhead. However, for some time-series sensory modalities, such as force and position, ${\bm L}_u$ can be larger than the raw data. Since ${\bm L}_u$ has the same dimension for all sensory modalities, it would be over compressed if we make its dimension smaller than that of the smallest raw sensory modality.

Note that, $f_{uni}(\cdot,\theta_A)$ is the module named Unimodal Encoder A in Fig.~\ref{fig:sys} and it will be trained together with the Multimodal Cross Attention A in the edge server which will be introduced next.  

\subsection{Multimodal Cross Attention}
\label{sec:mca}
\begin{figure}[t]
\centering
    \includegraphics[width=0.4\textwidth]{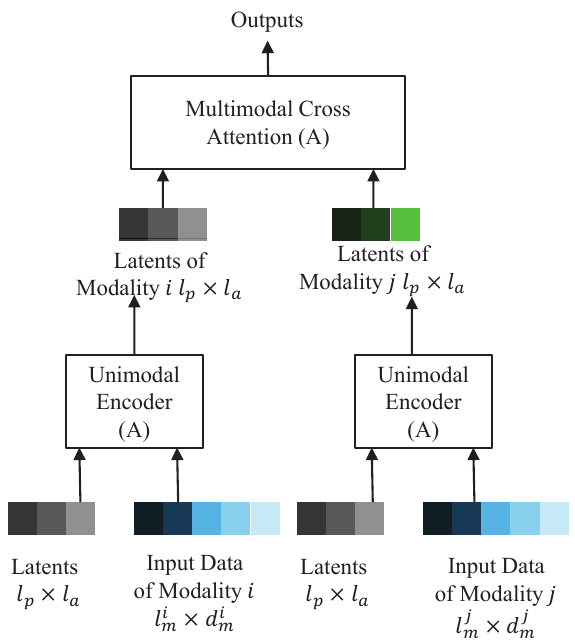}
    \vspace{-5pt}
    \caption{The HighMMT architecture in \cite{liang2022high} using unimodal encoder and multimodal cross attention.}
    \label{fig:highmmt}
    \vspace{-5pt}
\end{figure}

The edge server integrates latent data generated by Unimodal Encoder A in Fig.~\ref{fig:sys}. Following the multimodal integration approach proposed in \cite{liang2022high}, the Perceiver is used to integrate latent data as shown in Fig.~\ref{fig:highmmt}. If the $i$-th task requires $n_{t,i}$ modalities, then the model will run the multimodal cross attention module for $n_{t,i}(n_{t,i}-1)$ times to cover all combinations among the sensory modalities. The multimodal cross attention architecture is the same as the Perceiver used in the Unimodal Encoder A. Thus, the multimodal cross attention module can be modeled as:
\begin{align}
\label{equ:multicross}
    {\bm c}_{mul} = &f_{ffn}^{mul}(f_{mul}({\bm L}_1,{\bm L}_2;\theta_{A}^m),f_{mul}({\bm L}_1,{\bm L}_3;\theta_{A}^m),\cdots, \nonumber \\ 
    &f_{mul}({\bm L}_{n_{t,i}-1},{\bm L}_{n_{t,i}};\theta_{A}^m))
\end{align}
where $f_{ffn}^{mul}(\cdot)$ is a two-layer feedforward network with output size being the class number, $f_{mul}(\cdot;\theta_{A}^m)$ is the Perceiver architecture with parameter $\theta_A^m$. Then, using $\operatorname{argmax}({\bm c}_{mul})$, we can obtain the classification class label. 

Unimodal Encoder A will be trained together with Multimodal Cross Attention A using Cross Entropy Loss. It is the HighMMT model proposed in \cite{liang2022high} by connecting Unimodal Encoder A and Multimodal Cross Attention A. In this paper, it is named as Distributed-HighMMT since the two modules are in different devices.  

\subsection{Conformal Prediction in MuSeCo}
\label{sec:B}

To further reduce communication overhead, the AIoT device can perform additional processing to complete the task. For example, the AIoT device can send unimodal classification results directly to the edge server. At the edge server, the multimodal fusion process can adopt a majority vote approach, assuming each modality is equally important. However, sensory modalities often differ in data quality and dimensionality, making it inappropriate to treat them as equal.

The key idea behind MuSeCo is that the AIoT device uses unimodal Conformal Prediction \cite{angelopoulos2023conformal} to assess its confidence in the classification results. If the confidence is high, the sensory modality contains strong task-related information and can be fused in the edge server with higher weights. Conversely, if the confidence is low, the sensory modality lacks sufficient task-related information and may confuse the multimodal cross-attention module in the edge server. Therefore, confidence is a critical parameter generated by the Conformal Prediction process.

\begin{figure}[t]
\centering
    \includegraphics[width=0.47\textwidth]{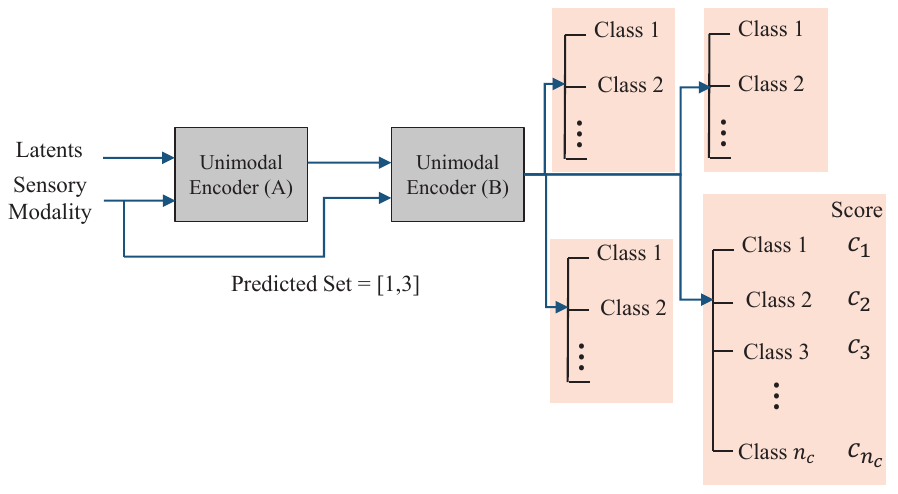}
    \vspace{-5pt}
    \caption{Illustration of the Conformal Prediction outputs for 4 tasks. Each task has its own output head to generate confidence scores and a prediction set.}
    \label{fig:conformal}
    \vspace{-5pt}
\end{figure}

To obtain Conformal Prediction results, the latent output of Unimodal Encoder A is fed into another unimodal encoder, along with the input sensory modality data. This can be modeled as:
\begin{align}
    {\bm l}_{cp} = f_{uni} ({\tilde {\bm M}}, {\bm L}_u; \theta_B), 
\end{align}
where $\theta_B$ is the parameter of the Unimodal Encoder B in Fig.~\ref{fig:sys}. Note that, the output ${\bm l}_{cp}$ is a vector which is the last column of the latent matrix. Then, a two-layer feedforward network $f_{ffn}^{uni}({\bm l}_{cp}; \theta_p, {\mathcal T}_i)$ is used, where $\theta_p$ is the parameter of the feedforward network and ${\mathcal T}_i$ is the task. Note that, this output head $f_{ffn}^{uni}({\bm l}_{cp}; \theta_p, {\mathcal T}_i)$ is task-dependent and we need one feedforward network for each task, as shown in Fig.~\ref{fig:conformal}. The first layer's input size is the length of ${\bm l}_{cp}$ and the second layer's output size is the class number of the dataset. In the end, Softmax function is used to generate scores as follows:
\begin{align}
    {\bm c} = \operatorname{softmax}({\bm l}_{cp}),
\end{align}
where ${\bm c} \in {\mathbb R}^{n_{c}}$ and $n_{c}$ is the class number of the task.   

By using $\operatorname{argmax}({\bm c})$ we can obtain the predicted class. The model can be trained using Cross Entropy Loss. Note that, Unimodal Encoder A is trained before Unimodal Encoder B and they are not trained together.

However, using $\operatorname{argmax}({\bm c})$ only provides a single predicted class without any indication of the model's confidence or certainty in its prediction. Conformal Prediction offers a solution by transforming the raw Softmax outputs ${\bm c}$ into meaningful scores that quantify prediction confidence. Intuitively, if the model is highly confident in its prediction, the largest value in ${\bm c}$ will be close to 1, while the other values will be significantly smaller. Conversely, if the model is uncertain, the largest value will be comparable to the second-largest value, indicating ambiguity in the prediction. Despite this, Softmax outputs alone do not represent true probabilities and cannot reliably quantify prediction confidence. Therefore, a transformation is necessary to convert these outputs into a form that accurately reflects the model's confidence.

To compute the Conformal Prediction score, a calibration dataset consisting of $n_{cal}$ data points ($x$,$y$) is required. This dataset must be distinct from the training data, where $x$ represents the input and $y \in \{1, 2, \cdots, n_{c}\}$  is the corresponding label. As described in \cite{angelopoulos2023conformal}, the Conformal Prediction score for a given data point is defined as ${c}_{cp} = 1-{\bm c}[y]$, where ${\bm c}[y]$ denotes the Softmax output corresponding to the true label $y$. By computing these scores for all $n_{cal}$ calibration examples, we obtain an array of scores ${\bm s}$. For a given $\alpha$, the quantile ${\hat q}$ is determined such that:
\begin{align}
    {\mathbb P}({c}_{cp}<{\hat q})\geq 1-\alpha, \text{ for } \forall {c}_{cp} \in {\bm s}.
\end{align}

This quantile ${\hat q}$ serves as a threshold for evaluating the confidence of predictions on new, unseen data. For a new input ($x$,$\cdot$) with an unknown label, the Softmax output vector ${\bm c}$ is compared against the threshold $1-{\hat q}$. Any class whose corresponding Softmax value exceeds this threshold is included in the prediction set ${\bm u}$. This approach ensures that the prediction set is constructed in a way that guarantees, with high probability, that the true label is included within the set. As demonstrated in \cite{angelopoulos2021uncertainty}, this method provides a statistically rigorous framework for quantifying confidence in predictions
\begin{align}
    {\mathbb P}(y \in {\bm u}) \approx 1-\alpha,
\end{align}
where $y$ is the true label. It is important to note that the size of the prediction set $|{\bm u}|$ is directly related to the model's confidence in its prediction. Specifically, a smaller set size indicates higher confidence, as the model is confident enough to narrow down the prediction to a few likely classes. Conversely, a larger set size reflects lower confidence, as the model is less confident and includes more classes in the prediction set to account for potential ambiguity.

Using Conformal Prediction, the AIoT device can transmit both the Softmax output vector 
${\bm c}$ and the prediction set ${\bm u}$ to the edge server. The edge server then integrates these outputs across multiple sensory modalities to generate a final, unified prediction. The goal is to combine the modalities in a way that accounts for their respective confidence levels, ensuring that more reliable modalities contribute more significantly to the final decision.

For a given task ${\mathcal T}$ involving multiple sensory modalities $\{{\bm M}_1, {\bm M}_2, \cdots \}$, the edge server receives the corresponding Softmax outputs $\{{\bm c}_1, {\bm c}_2, \cdots \}$ and associated prediction sets $\{{\bm u}_1, {\bm u}_2, \cdots \}$ from the AIoT devices. To combine these outputs effectively, we propose two distinct approaches: Equal-Weight Combination (EWC) and Set-Size-Scaled Combination (SSSC).
 
In EWC, all sensory modalities are considered equally. The Softmax outputs are added together, upon which the predict label ${\hat y}$ can be obtained as follows:
\begin{align}
\label{equ:ewc}
    {\bm c}_t = \sum_{i=1}^{|{\mathcal T}|} {\bm c}_i, \text{ and } {\hat y}={\operatorname{argmax}}({\bm c}_t),
\end{align}
where $|{\mathcal T}|$ is the number of sensory modalities in the task. In this way, even the sensory modality quality is low, it will be considered the same as the high-quality sensory modality. The certainty or confidence of each sensory modality generated by Conformal Prediction is not used in EWC.

To address this issue, we use the prediction set size as an indicator of confidence and scale the Softmax outputs. In SSSC, the modality combination is written as
\begin{align}
\label{equ:sssc}
    {\bm c}_t = \left(\sum_{m=1}^{|{\mathcal T}|} \frac{{\bm c}_m}{|{\bm u}_m|^{\beta}}\right)/\left(\sum_{m=1}^{|{\mathcal T}|}\frac{1}{|{\bm u}_m|^{\beta}}\right), \text{and } {\hat y}={\operatorname{argmax}}({\bm c}_t),
\end{align}
where $\beta \geq 1$ is a constant and $|{\bm u}|$ is the Conformal Prediction set size. When the task class number is small, e.g., binary classification with 2 classes, the prediction set size $|{\bm u}|$ is also small (no larger than 2) but this does not represent strong confidence. By using a large $\beta$, we can penalize large sets and reward small sets. For example, consider that $\beta=2$. For a set size 1, its Softmax outputs would not be changed, while for a set size 2, its Softmax outputs would be scaled by 0.25, which means higher uncertainty reduces its weight. Using SSSC, we can mitigate the impact of modality heterogeneity by reducing the low-quality sensory modality's weight in the combination stage. 

At this stage, we have introduced two approaches for task-oriented mulsemedia communication. First, we introduce the Distributed-HighMMT which uses the Unimodal Encoder A and Multimodal Cross Attention A, as shown in Fig.~\ref{fig:sys}. The AIoT devices send unimodal latent data to the edge server for multimodal fusion. Second, we introduce the MuSeCo model where the Unimodal Encoder A and B generate Conformal Prediction scores and sets which are sent to the edge server for weighted multimodal combination. In the next Section, we will introduce the MuSeCo-Adaptive which can combine the Distributed-HighMMT and MuSeCo. Also, we introduce the communication protocol and the evaluation parameters.

\section{Adaptive Mulsemedia Communication System}

MuSeCo effectively reduces data communication overhead by transmitting only Conformal Prediction scores and sets, though it typically achieves lower classification accuracy compared to Distributed-HighMMT, as demonstrated in Fig.~\ref{fig:no_noise_acc}. Unlike Distributed-HighMMT, which requires all sensory modalities to function and fails in their absence, MuSeCo is capable of making predictions with just a single sensory modality. Consequently, this presents a trade-off among classification accuracy, reliability, and data communication overhead. To address this and achieve a balanced solution, we introduce the MuSeCo-Adaptive.

\subsection{Adaptive Model Selection}

The basic idea of MuSeCo-Adaptive is to use the combined Softmax outputs ${\bm c}_t$ in Equ.~\eqref{equ:ewc} and Equ.~\eqref{equ:sssc}. If we rank the elements in ${\bm c}_t$ and the largest value is much larger than the second largest one, this means the modality combination has reached to a strong agreement. However, if the largest value is similar to the second largest one, the result of the modality combination is vague.  

\begin{figure}[t]
\centering
    \includegraphics[width=0.25\textwidth]{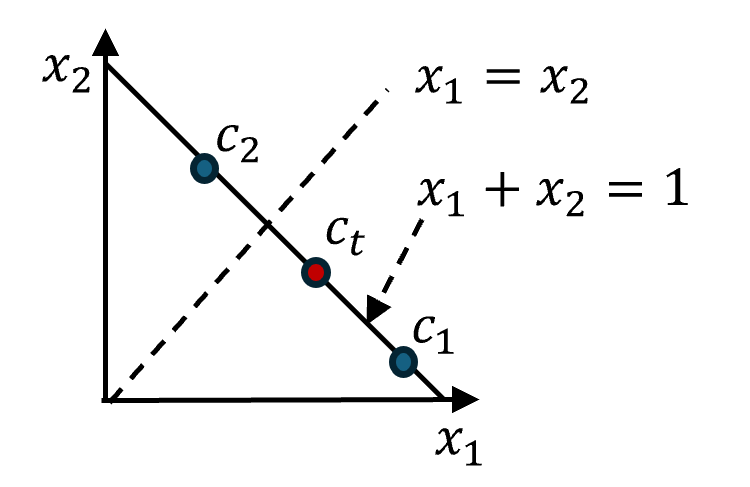}
    \vspace{-5pt}
    \caption{An example of the combined score ${\bm c}_t$ for a binary classification problem.}
    \label{fig:c_score}
    \vspace{-5pt}
\end{figure}

In Fig.~\ref{fig:c_score}, we use a toy example to show the idea. We consider a binary classification problem with two sensory modalities. The unimodal Softmax outputs are ${\bm c}_1=[x_1^1,x_2^1]$ and ${\bm c}_2=[x_1^{2}, x_2^2]$. If we use the EWC in Equ.~\eqref{equ:ewc}, we have ${\bm c}_t=({\bm c}_1+{\bm c}_2)/2$. In Fig.~\ref{fig:c_score}, the line $x_1=x_2$ is a boundary. If $c_t$ is on the left-hand side, the result prefers class 2, and vice versa. The ${\bm c}_1$, ${\bm c}_2$, and ${\bm c}_t$ must be on the line $x_1+x_2=1$.  

We examine the following three scenarios to demonstrate the behavior of the system:
\begin{itemize}
    \item {\bf Strong Agreement:} If ${\bm c}_1$ is close to [1,0] and ${\bm c}_2$ is on the right-hand side of the boundary, ${\bm c}_t$ will be close to [1,0] and the two sensory modalities reach to an agreement. Using EWC or SSSC, we can obtain the same prediction result. 
\item
    {\bf Contradictory Evidence:} If ${\bm c}_1$ is close to [1,0] and ${\bm c}_2$ is close to [0,1], ${\bm c}_t$ will be close to $x_1=x_2$. It is challenging to predict the label since the two sensory modalities provide contradict evidence.
\item
    {\bf Ambiguity Near the Boundary:} If both ${\bm c}_1$ and ${\bm c}_2$ are close to the boundary $x_1=x_2$, the model cannot make a strong decision. 
\end{itemize}

Generally, if $\max({\bm c}_t)$ is smaller than a threshold, the approach based on Softmax outputs and Conformal Prediction is not efficient and we should use the Multimodal Cross Attention module. To determine the optimal threshold, we reuse the calibration dataset that was used for Conformal Prediction.

\begin{figure*}[t]
     \centering
     \begin{subfigure}[b]{0.245\textwidth}
         \centering
         \includegraphics[width=\textwidth]{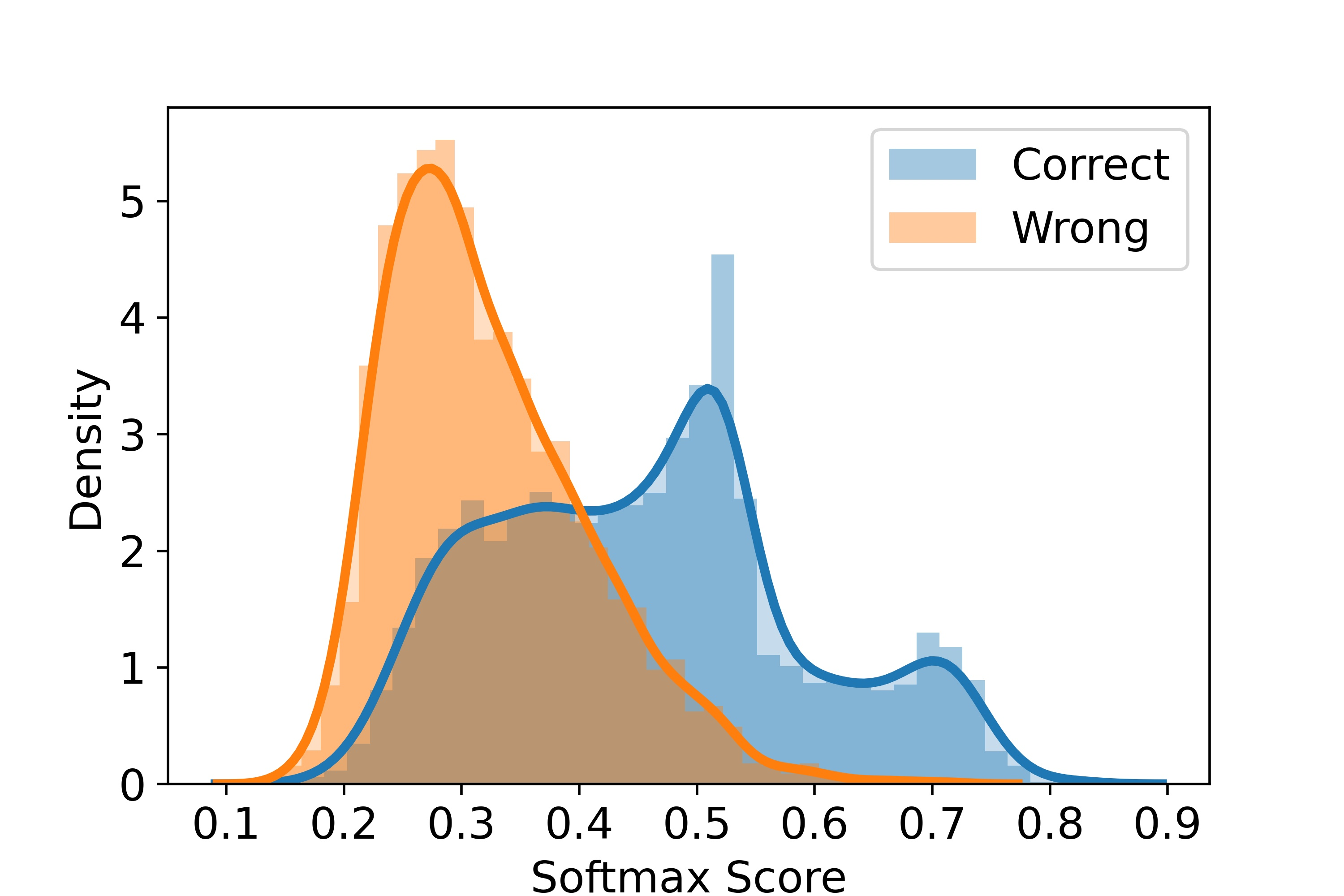}
         \caption{Task 1.}
         \label{fig:distribution_of_cscore_ewc}
     \end{subfigure}
     \hfill
     \begin{subfigure}[b]{0.245\textwidth}
         \centering
         \includegraphics[width=\textwidth]{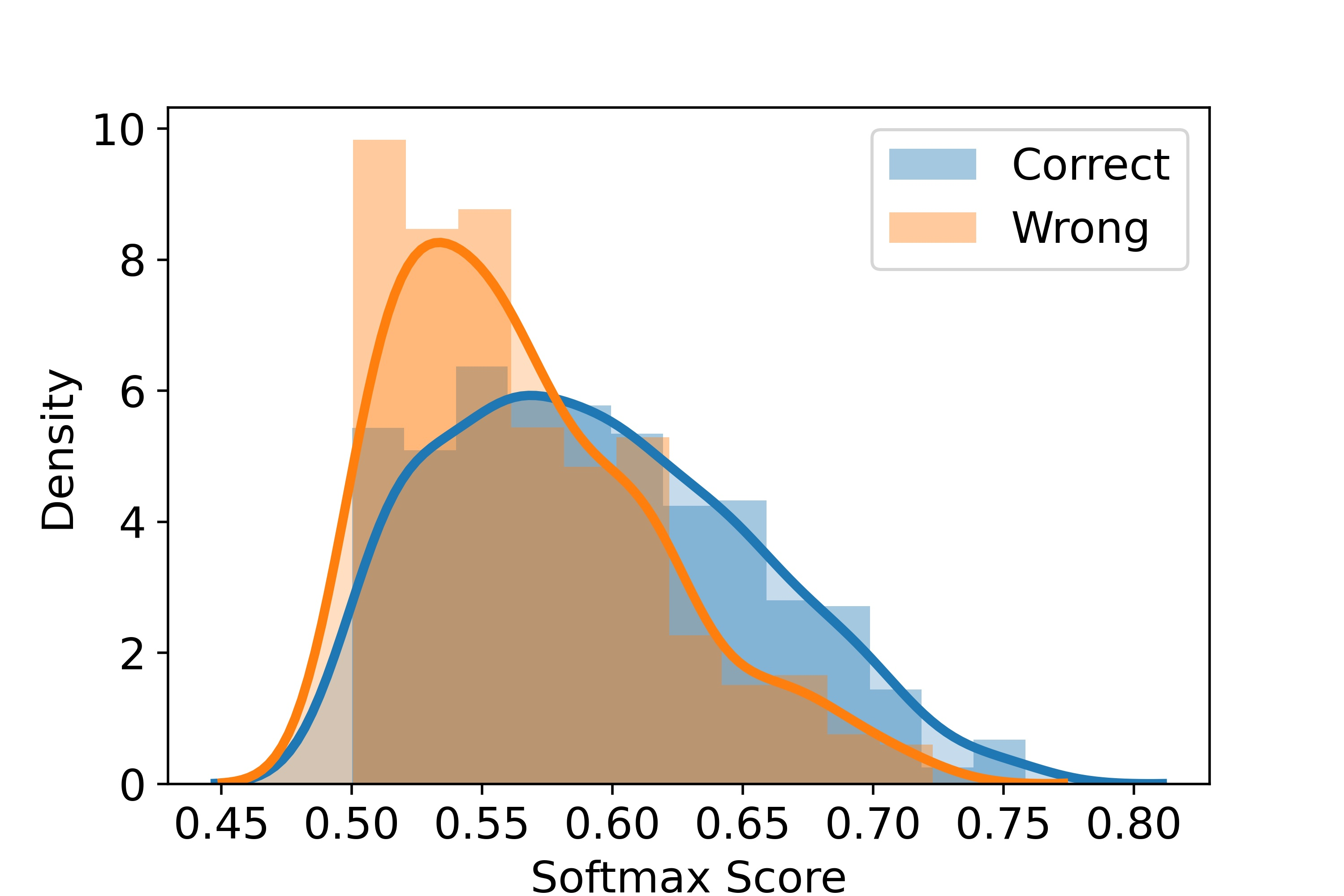}
         \caption{Task 2.}
         \label{fig:distribution_of_cscore_ewc_2}
     \end{subfigure}
     \hfill
     \begin{subfigure}[b]{0.245\textwidth}
         \centering
         \includegraphics[width=\textwidth]{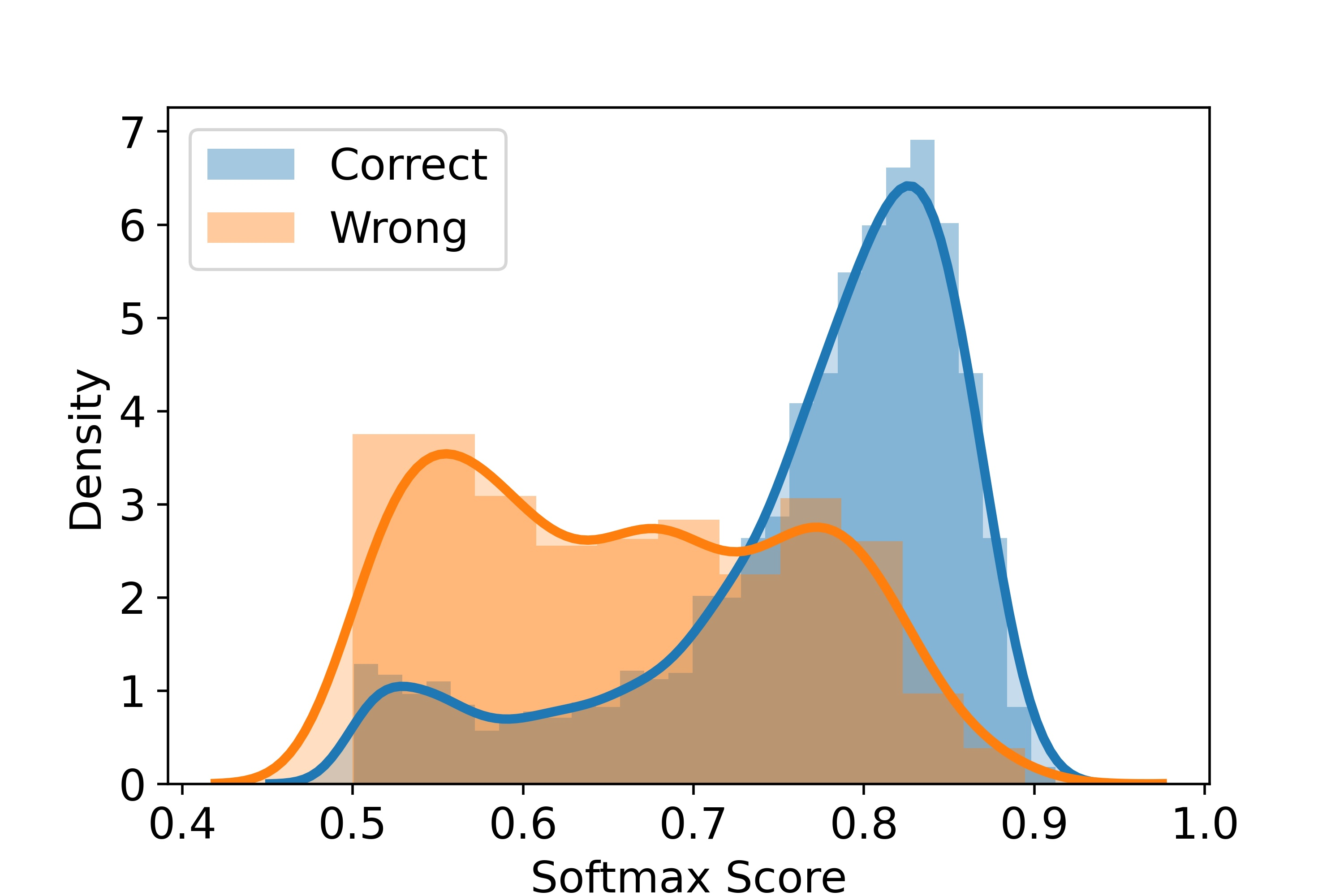}
         \caption{Task 3.}
         \label{fig:distribution_of_cscore_ewc_3}
     \end{subfigure}
          \hfill
     \begin{subfigure}[b]{0.245\textwidth}
         \centering
         \includegraphics[width=\textwidth]{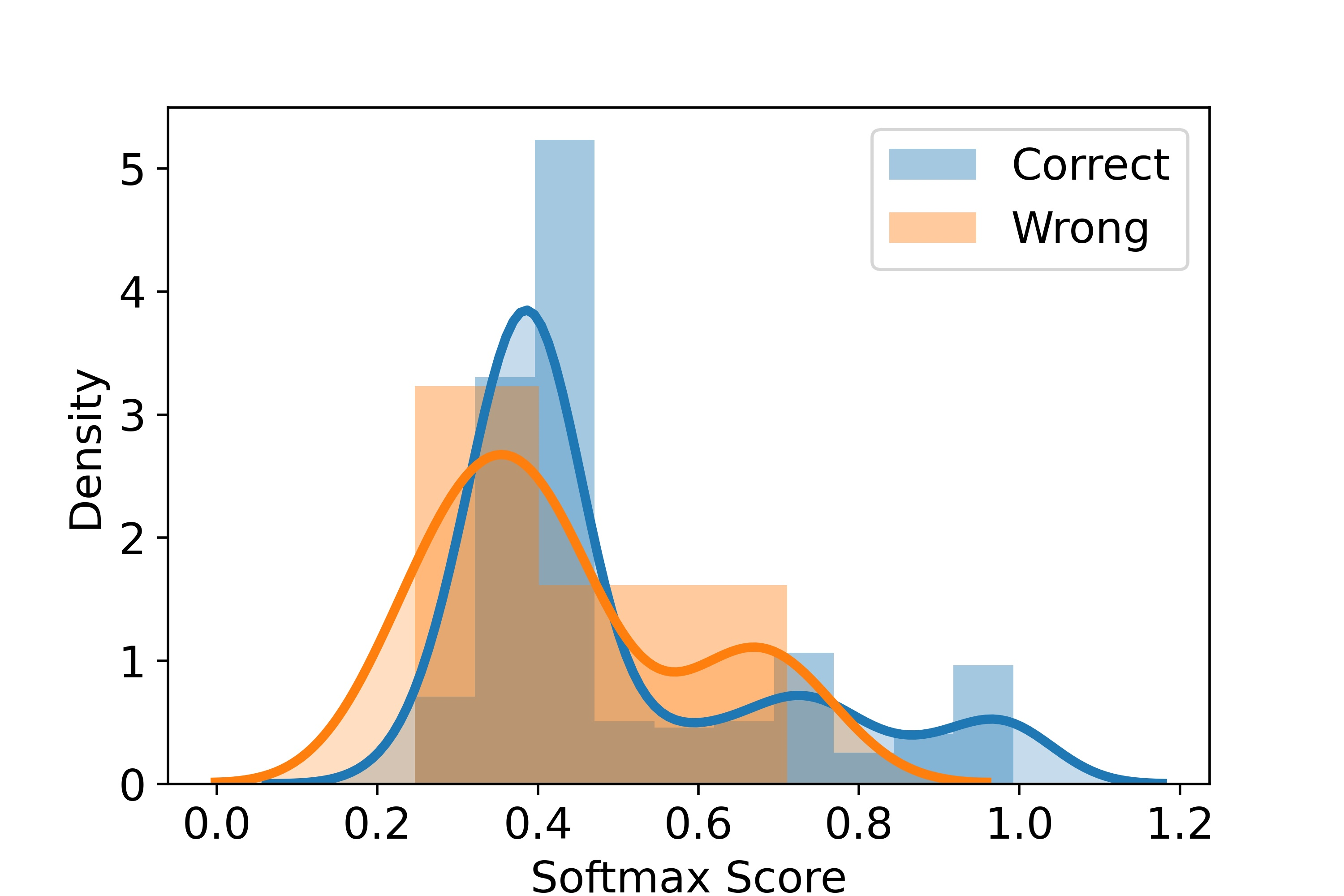}
         \caption{Task 4.}
         \label{fig:distribution_of_cscore_ewc_4}
     \end{subfigure}
        \caption{Softmax score distribution of equal-weight combination (EWC) for correctly and wrongly classified samples.}
        \label{fig:ewc}
\end{figure*}
\begin{figure*}[t]
     \centering
     \begin{subfigure}[b]{0.245\textwidth}
         \centering
         \includegraphics[width=\textwidth]{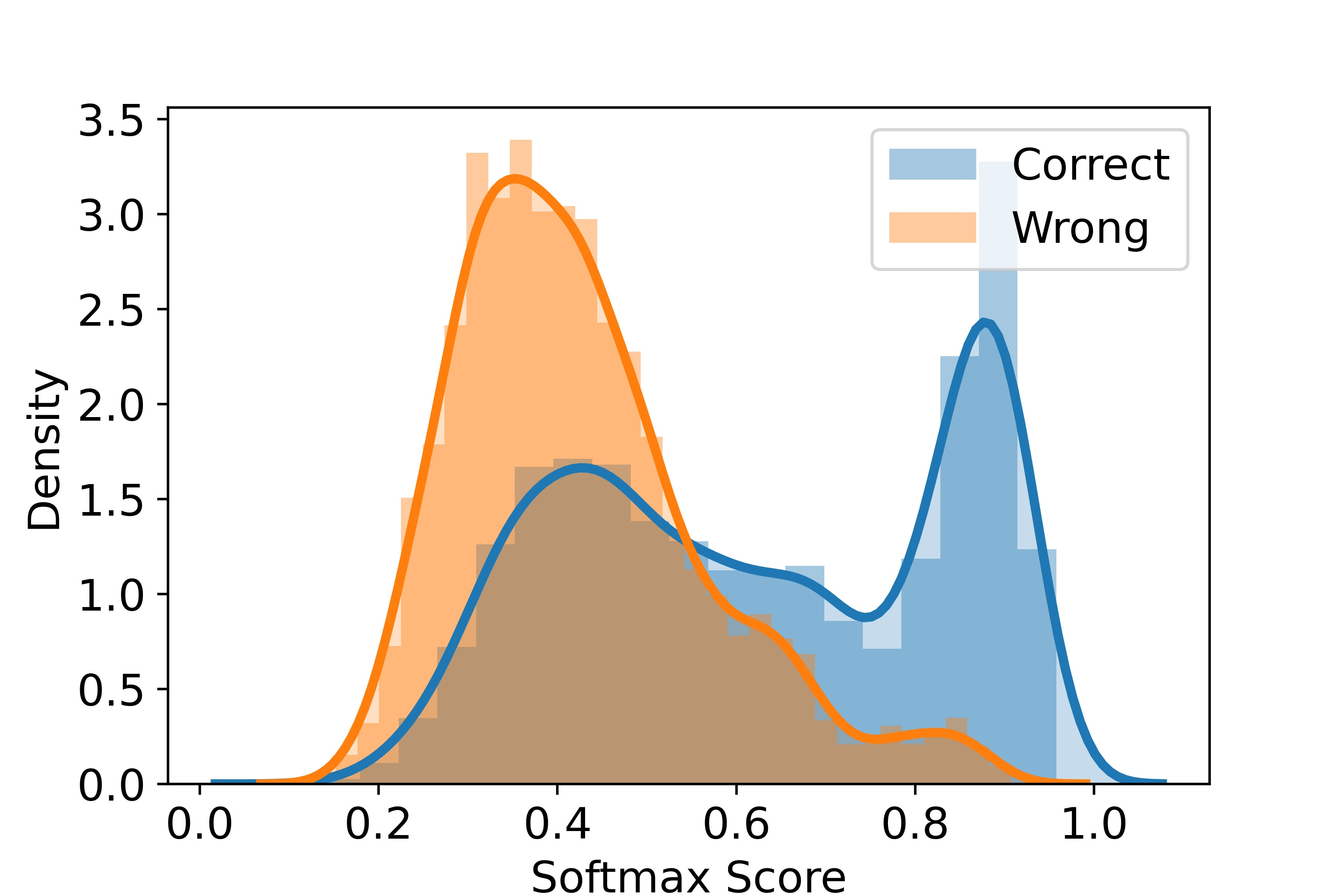}
         \caption{Task 1. }
         \label{fig:distribution_of_cscore_sssc}
     \end{subfigure}
     \hfill
     \begin{subfigure}[b]{0.245\textwidth}
         \centering
         \includegraphics[width=\textwidth]{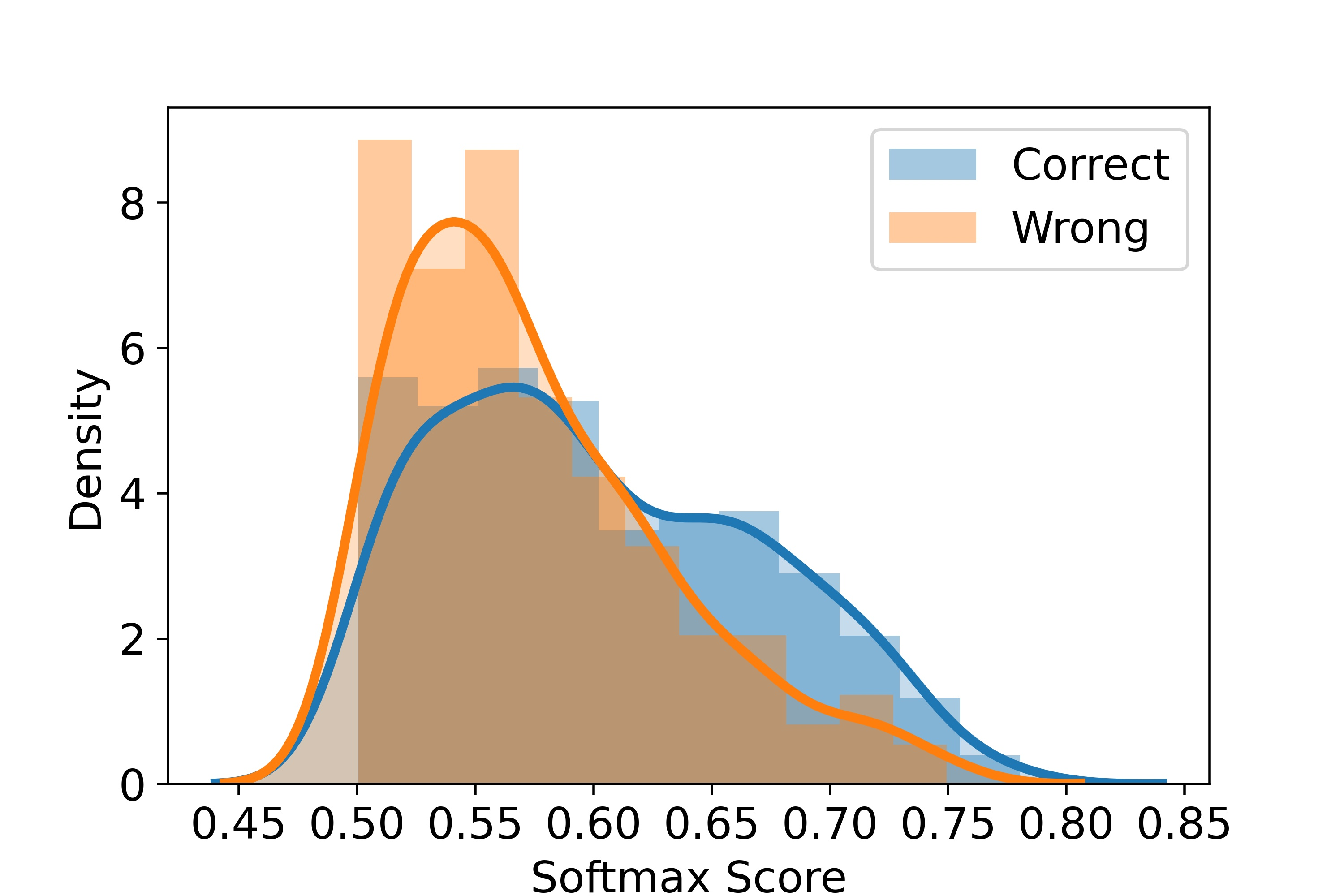}
         \caption{Task 2.}
         \label{fig:distribution_of_cscore_sssc_2}
     \end{subfigure}
     \hfill
     \begin{subfigure}[b]{0.245\textwidth}
         \centering
         \includegraphics[width=\textwidth]{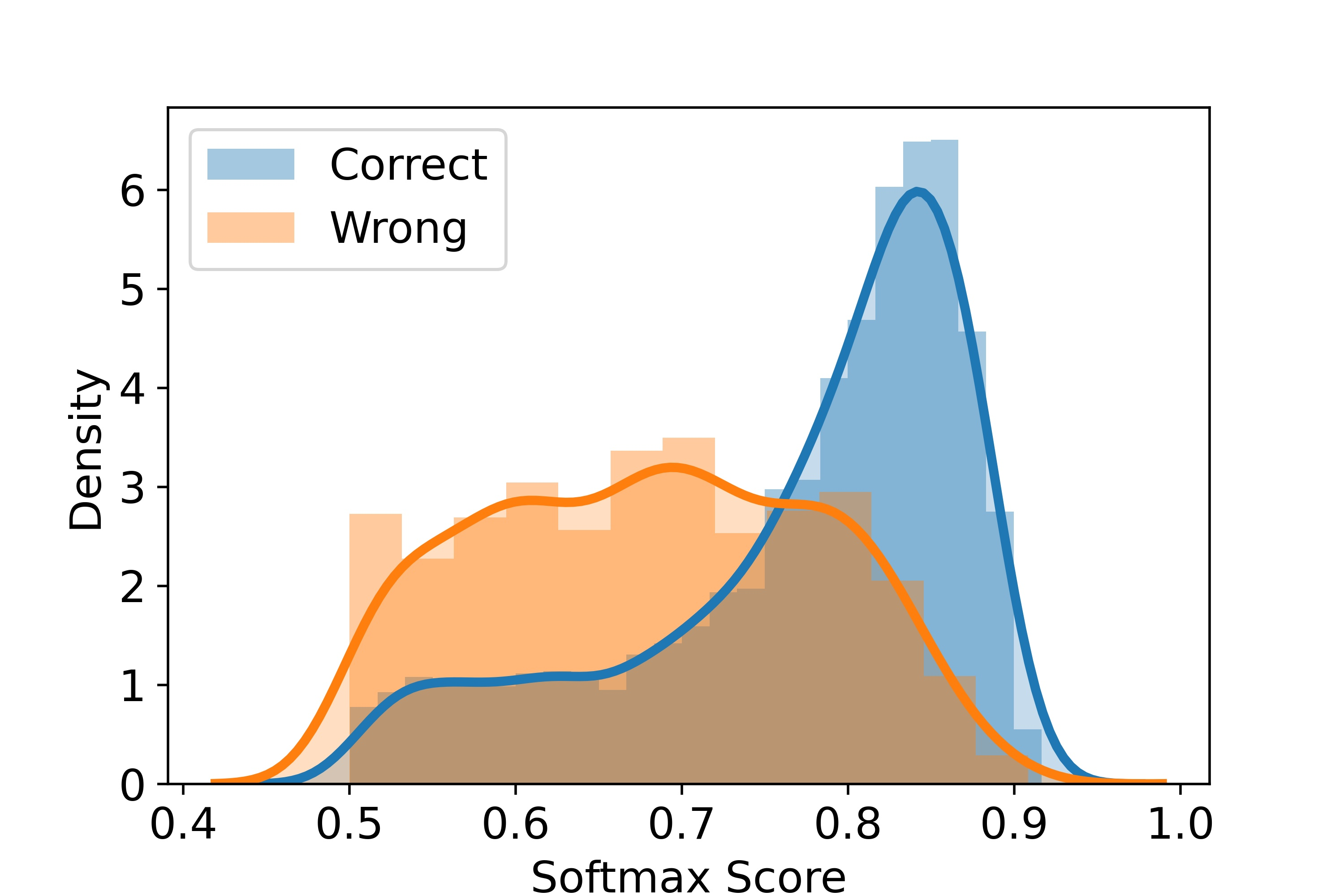}
         \caption{Task 3.}
         \label{fig:distribution_of_cscore_sssc_3}
     \end{subfigure}
          \hfill
     \begin{subfigure}[b]{0.245\textwidth}
         \centering
         \includegraphics[width=\textwidth]{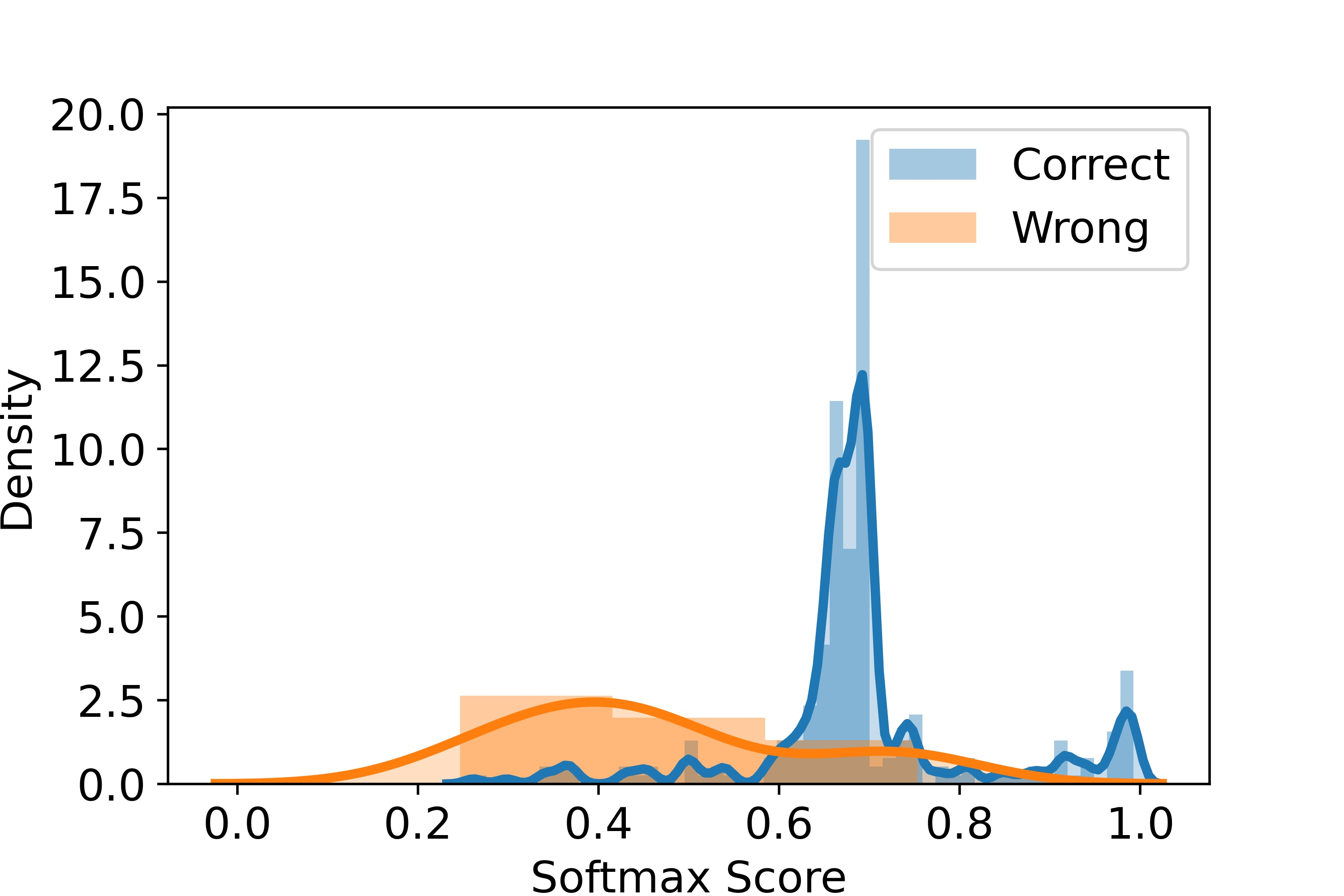}
         \caption{Task 4.}
         \label{fig:distribution_of_cscore_sssc_4}
     \end{subfigure}
        \caption{Softmax score distribution of set-size-scaled combination (SSSC) for correctly and wrongly classified samples.}
        \label{fig:sssc}
\end{figure*}

In Fig.~\ref{fig:ewc} and Fig.~\ref{fig:sssc} we show the density of Softmax score for test samples with correct and wrong labels using EWC and SSSC, respectively. We use MuSeCo and evaluate 4 tasks given in Table~\ref{tab:modality}. As discussed before, AIoT devices send Softmax outputs and Conformal Prediction sets to the edge server. We consider the data can be precisely received without any error. Then, the edge server uses EWC and SSSC with $\beta=1$ to combine sensory modalities. As we can see in Fig.~\ref{fig:ewc}, the majority of the correctly classified samples have higher Softmax scores than the misclassified samples in Task 1, Task 3, and Task 4 using SSSC. In Table~\ref{tab:score_compare}, we provide a comprehensive comparison of the results in Fig.~\ref{fig:ewc} and Fig.~\ref{fig:sssc}. We show the 30\% and 70\% percentile of the classified samples with correct and wrong labels, respectively. For example, the 30\% percentile of the classified samples with correct labels for Task 1 using EWC is 0.37 which means 30\% of correctly classified samples' combined Softmax scores are smaller than 0.37, while 70\% of the misclassified samples' softmax scores are smaller than 0.36, which shows that the combined Softmax scores of the correct and wrong groups have different distributions. A large difference between the two percentiles indicates that the combined Softmax scores of correct and wrong samples are more different. Moreover, if we use the 30\% percentile of the correctly classified samples as a threshold to obtain a subset of the original dataset, in which all samples have the maximum Softmax score larger than the threshold. In this subset, we can obtain the accuracy and the covered sample percentage in Table~\ref{tab:score_compare}. We can consider these samples are simple ones and those below the threshold are complex ones. As we can see from the table, for all the four tasks, around more than 57\% samples's maximum Softmax scores are above the threshold and the obtained accuracy is higher than using the complex Multimodal Cross Attention module (the SOTA in Table~\ref{tab:modality}).

\begin{table}[t]
\centering
\caption{Comparison between EWC and SSSC based on the Softmax score distribution.}
\begin{tabular}{|l|c|p{1cm}|p{1cm}|p{1.2cm}|p{1.2cm}|}
\hline
\multicolumn{2}{|c|}{{Tasks}}  & Correct (30\%) & Wrong (70\%)  & Accuracy (\%)& Sample Percentage (\%)\\
\hline

\multirow{2}{*}{Task 1} & EWC & 0.37 & 0.36  &  86.02 &  56.92\\
                               & SSSC & 0.47 & 0.47   &  82.19 &56.84\\
            \hline
\multirow{2}{*}{Task 2} & EWC & 0.56 & 0.59   &  71.90 & 59.57 \\
                               & SSSC & 0.55 & 0.60   &  69.10 &68.59\\
            \hline
\multirow{2}{*}{Task 3} & EWC & 0.75 & 0.73   &  88.47 & 58.17 \\
                               & SSSC & 0.75 & 0.74   &  88.49 &59.74\\
            \hline
\multirow{2}{*}{Task 4} & EWC & 0.40 & 0.44  & 97.85 &  68.38 \\
                               & SSSC & 0.66 & 0.49   &  98.92 &68.38\\
            \hline
\end{tabular}
\label{tab:score_compare}
\end{table}


It should be noted that, the numbers of correctly classified and misclassified samples are different. Although the distributions for Task 2 are similar, the accuracy remains high. Also, we notice that EWC and SSSC have similar performance in Tasks 1 to 4. Since here we do not consider noises in sensory modality, there is no significant differences in the predictions sets. 

Based on the intuitive insights and the empirical analysis, we have shown that a percentile can be used to obtain a threshold. As we can see in Table~\ref{tab:score_compare}, the 30\% percentile varies across different tasks and combination approaches. To address this issue, we use calibration datasets for each task to obtain the threshold percentile. First, by using the calibration datasets $\{x_{cal},y_{cal}\}$ and a percentage $\alpha_2$, we obtain a quantile for each task ${\hat q}_i^e$ by only considering the Softmax score of the correctly predicted samples. Note that, this quantile is used in the edge server, which is different from the quantile used in Conformal Prediction. Above the threshold, we consider the sensory modalities have consistent semantic meaning and the combined Softmax score will be used to generate the prediction outputs, while below the threshold, the Multimodal Cross Attention A module is used. 

\subsection{Mulsemedia Communication Protocol}
\label{sec:protocol}
\begin{figure}[t]
\centering
    \includegraphics[width=0.48\textwidth]{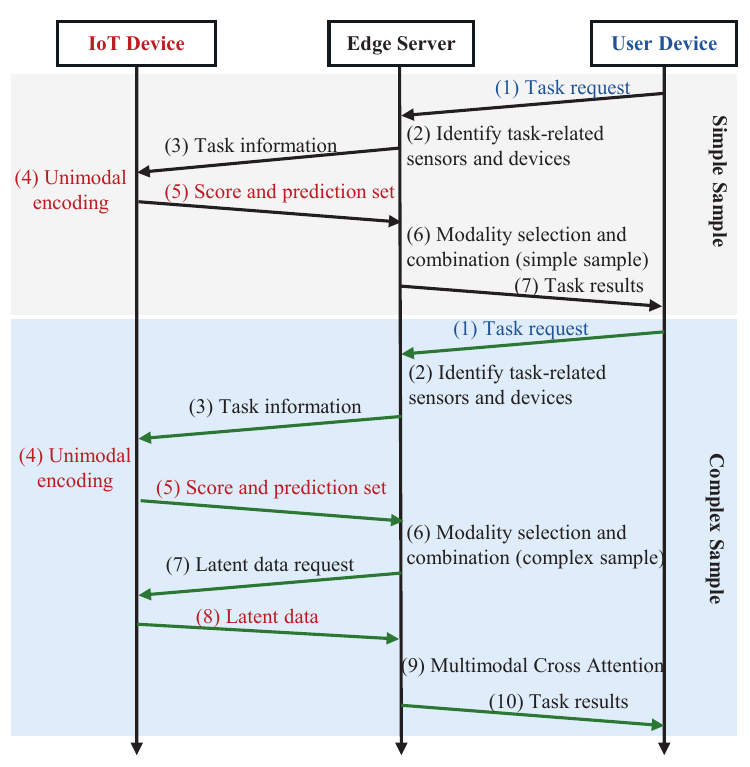}
    \vspace{-5pt}
    \caption{Illustration of MuSeCo-Adaptive communication protocol. }
    \label{fig:protocol}
    \vspace{-5pt}
\end{figure}

Based on previous analyses, we present the MuSeCo-Adaptive communication protocol, which is also shown in Fig.~\ref{fig:protocol}. The user device can request for task results. The edge server identifies the task-related sensory modalities, sensors, and devices and then send requests to identified AIoT devices. Each selected AIoT device collects sensory data and use Unimodal Encoder A and B to generate Softmax scores and Conformal Prediction set which are sent to the edge server. If the combined maximum score is higher than the threshold $\alpha_2$, the sample is considered simple and the edge server will send the results to the user device and complete the process.

If the combined maximum score is lower than the threshold $\alpha_2$, the sample is considered complex. It is challenging to classify this sample using EWC or SSSC. The edge server will request for more latent data from the Unimodal Encoder A. Since the latent data was used to generate Conformal Prediction sets, AIoT devices do not need to recompute and they can send saved latent data. After receiving the latent data, the edge server uses the Multimodal Cross Attention A module to generate the final classification result which will be sent to the user device.    

In order to comprehensively compare the performance of MuSeCo and MuSeCo-Adaptive, we introduce five communication approaches with various setups. The five approaches include: 
\begin{itemize}
    \item {\bf A1 (Raw Data)}: The AIoT devices send raw sensory data to the edge server. The edge server uses HighMMT model \cite{liang2022high} to process the sensory data and generate classification results. 
    \item {\bf A2 (Task Results)}: Each AIoT device uses Unimodal Encoder A and B to generate classification results and sends them to the edge server. The edge server performs a majority vote (randomly select when the sensory modality number is even) and send results to user device.
    \item {\bf A3 (Latent Data)}: AIoT devices send latent data output from Unimodal Encoder A to the edge server, where Multimodal Cross Attention A module is used to generate the task outputs which are sent to user devices. 
    \item {\bf A4 (MuSeCo: Softmax Scores and Conformal Prediction Sets)}: AIoT devices send Softmax scores and Conformal Prediction sets to the edge server. The edge server uses SSSC to select and combine sensory modalities to generate task outputs and send to user devices. 
    \item {\bf A5 (MuSeCo-Adaptive: Adaptive Transmission)}: AIoT devices send Softmax scores and Conformal Prediction sets to the edge server. If the edge server decides that the sample is a complex one, it requests for latent data from each AIoT device. Then, it uses Multimodal Cross Attention A module to generate the task outputs. The details are shown in Fig.~\ref{fig:protocol}. 
\end{itemize}

\begin{figure*}[t]
\centering
    \includegraphics[width=0.8\textwidth]{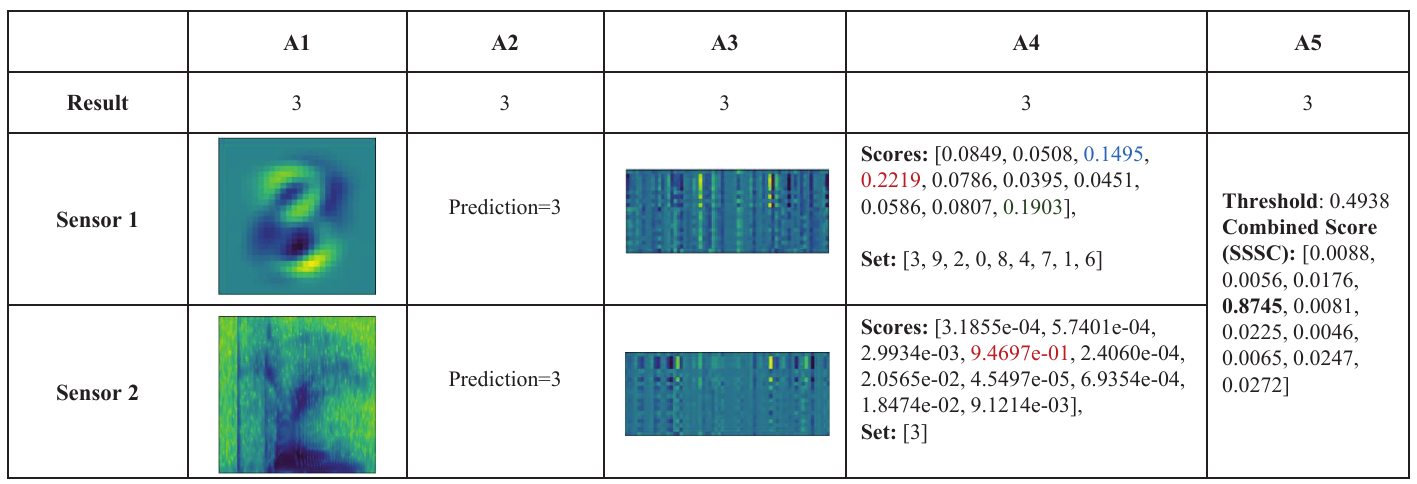}
    \vspace{-1pt}
    \caption{ An example of the approaches A1 to A5 for Task 1 (AV-MNIST).}
    \label{fig:example}
    \vspace{-5pt}
\end{figure*}
An example of the 5 approaches for Task 1 is given in Fig.~\ref{fig:example}. There are two AIoT devices and they are equipped with image and audio sensors, respectively. In A1, the AIoT devices send image and audio spectrogram to the edge server which uses the HighMMT model to identify the digit. In A2, each AIoT device sends its classification result to the edge server which will select the most common result. In A3, each AIoT device sends its latent data to the edge server, and the edge server uses Multimodal Cross Attention A module to generate the classification result. In A4, each AIoT device sends the unimodal Softmax score and Conformal Prediction sets with 90\% confidence to the edge server. For example, Sensor 1 has a high uncertainty and the largest three Softmax scores are close which are associated with the numbers 3, 9, and 2. With 90\% confidence, the true digit is within the set [3,9,2,0,8,4,7,1,6]. The large set size indicates that Sensor 1's result is not reliable. On the contrary, Sensor 2's set only has number 3 which indicates high confidence. In SSSC combination, Sensor 1's Softmax score will be discounted by $1/9^\beta$ and then added with Sensor 2's Softmax score. In A5, each AIoT device sends the same Conformal Prediction information as that in A4 to the edge server. The calibration threshold is 0.4983. If the combined maximum score (e.g., using SSSC) is higher than 0.4983, the edge server will proceed to make a prediction. In this example, the maximum combined score is 0.8745 which is higher than the threshold and the result is digit 3. However, if the maximum combined score is lower than 0.4983, the edge server will request for latent data and the Multimodal Cross Attention A module will be used.      

Next, we introduce the communication system and evaluation parameters, which are followed by comprehensive comparisons in the next section.

\subsection{Communication and Evaluation Parameters}
In this paper, we evaluate the trade-offs among the task accuracy, end-to-end latency, and energy consumption. In order to simulate a wireless communication system with various signal-to-noise ratios (SNR), we convert float decimal data into binary data and then send the binary data using modulated symbols. We consider uncoded modulation schemes and each float decimal number is encoded into 18 bits with 9 bits for the integer part and 9 bits for the fraction part. This customized encoding approach is used to efficiently use the binary bits. More universal ways are also available using typical floating number standards. As we can see in Fig.~\ref{fig:example}, the data transmitted from the sensor device is drastically different using different approaches. {For example, in A1 the transmitted raw sensory data is 240 kbits and in A3 the transmitted latent data is 23 kbits, while it is only 36 bits in A2. We consider the raw sensory data and latent data are transmitted using 64QAM modulation due to their large data size, while classification results, Softmax scores, and Conformal Prediction sets are sent using QPSK which can be reliably received. AIoT devices use the same transmission power $P_t$ for data communication.} 

We assume all AIoT devices have the same computing capacity which is represented by $c_{iot}$. The computing capacity is measured using Floating-point Operations Per Second (FLOPS).
The edge server's computing capacity is $c_s$ and one Perceiver unit (i.e., Unimodal Encoder A and B, the Multimodal Cross Attention A) requires $T_b$ floating-point operations. 

The overall required computation for task $i$ is related to the number of modalities and the communication approaches. For A1 the total computation is 
\begin{align}
    T_1 \approx T_b(|{\mathcal {T}}|)+T_b(|{\mathcal {T}}|-1)|{\mathcal {T}}|.
\end{align}

The end-to-end latency is defined as the time from the user device requests for task output to the time it receives the result. Since the majority of the latency consists of the transmission of raw data from AIoT devices and the computation in the edge server, the end-to-end latency for a task $i$ is approximated by
\begin{align}
    \tau_1 = \sum_{m=1}^{|{\mathcal {T}}|} |{\bm M}_m|/r_t+T_1/c_s.
\end{align}
where $r_t$ is the data rate. The AIoT devices use Time Division Multiple Access (TDMA) and the data transmission is sequential.

Similarly, the energy consumption for task $i$ is approximated using the communication and computation energy consumption, which is
\begin{align}
    E_1 \approx P_t\times(\sum_{m=1}^{|{\mathcal {T}}|} |{\bm M}_m|/r_t)+T_1/\Gamma_s,
\end{align}
where $\Gamma_s$ is the energy efficiency of the server, which is measured in FLOPS/W (FLOPs/Joule) \cite{DESISLAVOV2023100857}. The task accuracy is measured based on the ML model outputs and the ground truth labels.

In A2, the majority of the computation task is performed in the AIoT device and the latency is
\begin{align}
    \tau_2 \approx 2T_b/c_{iot}.
\end{align}
Here, we implicitly assume that all the AIoT devices' computations are parallel. The total energy consumption can be written as
\begin{align}
    E_2  \approx  2T_b(|{\mathcal {T}}|)/\Gamma_{iot},
\end{align}
where $\Gamma_{iot}$ is the energy efficiency of the AIoT device.
The communication latency and communication energy consumption are neglected since each AIoT device only sends its classification task results.

In A3, the latency consists of three major parts, i.e., the computation in the AIoT device, the latent data transmission from the AIoT device to the edge server, and the computation in the edge server. Thus, we have
\begin{align}
    \tau_3 \approx T_b/c_{iot}+\sum_{m=1}^{|{\mathcal {T}}|} |{\bm L}_{u,m}|/r_t+(|{\mathcal {T}}|-1)(|{\mathcal {T}}|)T_b/c_s,
\end{align}
where ${\bm L}_{u,m}$ is the latent data of the $m$-th sensory modality for a task.
The total energy consumption can be written as
\begin{align}
    E_3 \approx T_b\cdot|{\mathcal {T}}|/\Gamma_{iot}+\sum_{m=1}^{|{\mathcal {T}}|} |{\bm L}_{u,m}|P_t/r_t+|{\mathcal {T}}|(|{\mathcal {T}}|-1) T_b/\Gamma_{s}.
\end{align}

In A4, the latency $\tau_4$ and energy consumption $E_4$ can be approximated using $\tau_2$ and $E_2$, respectively. The Softmax scores and Conformal Prediction sets have limited data and the computation is mainly performed in AIoT devices. 

In A5, the latency and energy consumption depend on the percentage of simple samples, which is denoted by $p_{h}$. For simple samples, the latency is $\tau_4$ and the energy consumption is $E_4$. For complex samples, the latency consists of three major parts: the unimodal encoding using Unimodal Encoder A and B, latent data transmission, and Multimodal Cross Attention computation. This is because the MuSeCo-Adaptive uses MuSeCo first and then uses latent data, as shown in Fig.~\ref{fig:protocol}. 
Here, we implicitly assume the steps (5) to (7) in Fig.~\ref{fig:protocol} take negligible time. As a result, the latency of A5 is
\begin{align}
    \tau_5 &\approx p_h(2T_b/c_{iot})+(1-p_h)(2T_b/c_{iot}+\sum_{m=1}^{|{\mathcal {T}}|} |{\bm L}_{u,m}|/r_t \nonumber \\
    &+(|{\mathcal {T}}|-1)(|{\mathcal {T}}|)T_b/c_s),
\end{align}
and the total energy consumption is
\begin{align}
    E_5 &\approx p_h (2T_b(|{\mathcal {T}}|)/\Gamma_{iot})+(1-p_h)(2T_b(|{\mathcal {T}}|)/\Gamma_{iot} \nonumber \\
    &+\sum_{m=1}^{|{\mathcal {T}}|} |{\bm L}_{u,m}|P_t/r_t+(|{\mathcal {T}}|-1)(|{\mathcal {T}}|)T_b/\Gamma_s).
\end{align}
The above latency and energy consumption models can be applied to any task. The only difference is the task sensory modality number and input sensory modality data volume.

\section{Mulsemedia Communication Performance Evaluation}
\label{sec:eval}

To evaluate the performance of the proposed models, we conducted experiments using four distinct tasks: Task 1 (AV-MNIST) \cite{vielzeuf2018centralnet}, Task 2 (UR-FUNNY) \cite{hasan-etal-2019-ur}, Task 3 (MOSEI) \cite{zadeh2018multimodal}, and Task 4 (modified PUSH) \cite{lee2020multimodal}. Table~\ref{tab:modality} provides an overview of the sensory modality dimensions and the SOTA classification accuracy for each task.

Task 1, AV-MNIST, involves the identification of digits (0–9) using both image and audio spectrogram data. This task is based on a modified version of the original MNIST dataset, where image quality is intentionally degraded, and additional noise is introduced into the audio component to create a more challenging multimodal classification problem. Tasks 2 and 3 focus on humor detection and emotion classification, respectively, leveraging multimodal data to achieve high accuracy. Task 4 is derived from the MUJOCO PUSH task \cite{lee2020multimodal}, which originally involves predicting the 2D location of an object manipulated by a robotic arm. To align with the classification framework of Tasks 1 to 3, we transformed Task 4 into a classification problem by dividing the surface into nine equal-sized regions. The prediction problem becomes a classification problem of identifying the region where the object will ultimately reside. Due to memory constraints, the dataset size for Task 4 was reduced.

{These tasks are important in mulsemedia communication. For example, emotion sensing in XR can enable streaming of customized interactive content for users. The HMD integrates various sensors including hand-tracking and smell sensors and displays \cite{akyildiz2023mulsemedia}. The multimodal sensors can collect data and semantic latent data can be obtained on the HMD. Then, the edge server can perform emotion sensing and customize the content that will be delivered to the user. Also, the system can be extended to other sensory modalities such as smell and taste. In XR-enabled food quality examination, a user can provide smell and taste data of food using sensors \cite{akyildiz2023mulsemedia}. A local microcomputer, such as cell phone, can obtain semantic latent data which is sent to edge servers. The server can provide food quality result which is sent back to the user.}

{We import the dataset from the MultiBench dataset \cite{liang2023multizoo}.} Our simulation and experiments are performed using NVIDIA V100 GPU with 32 GB memory. We train Distributed-HighMMT and MuSeCo each for 3 times and test their performance. The learning rate is 0.001 and the weight decay is 0.001. The depth of each Perceiver module is 1, the number of latents is 20, the latent dimension is 64, the number of heads for cross attention is 1, the number of heads for latent self attention is 6, and the cross attention head dimension and latent self attention head dimension are both 64. The models are offline trained and deployed for online execution.

\begin{table*}[htb]
\centering
\caption{Task sensory modality, input dimension and state of the art accuracy.} 
\begin{tabular}{|p{1.05cm}|l|l|l|l|l|l|l|l|p{1.05cm}|p{1.05cm}|l|l|}
\hline
  & \multicolumn{2}{c|}{{Task 1}} & \multicolumn{3}{c|}{{Task 2}}  & \multicolumn{3}{c|}{{Task 3}} & \multicolumn{4}{c|}{{Task 4}}\\
\hline
Sensory Modality&Image &Audio&Image&Audio&Text&Image&Audio&Text&Gripper Pos&Gripper Sensors&Image&Control\\
\hline
Dimension&28$\times$28&112$\times$112&20$\times$371&20$\times$81&50$\times$300&50$\times$35&50$\times$74&50$\times$300&16$\times$3&16$\times$7&16$\times$32$\times$32&16$\times$7\\ \hline
SOTA& \multicolumn{2}{c|}{{71.1}} & \multicolumn{3}{c|}{{66.2}}  & \multicolumn{3}{c|}{{80.2}} & \multicolumn{4}{c|}{{---}}\\
            \hline
\multicolumn{13}{l}{\footnotesize{*The SOTA accuracy is obtained from \cite{liang2022high}. Task 1 is the AV-MNIST \cite{vielzeuf2018centralnet}, Task 2 is the UR-FUNNY \cite{hasan-etal-2019-ur}, Task 3 is the MOSEI \cite{zadeh2018multimodal}, and Task 4 is}}\\
\multicolumn{13}{l}{\footnotesize{modified PUSH \cite{lee2020multimodal}.}}\\
\end{tabular}

\label{tab:modality}
\end{table*}

\subsection{Reliable Communication}

The development of 5G and Beyond wireless systems supports ultra-reliable communication for AIoT systems, particularly the Ultra-Reliable Low Latency Communication (URLLC) service. In this subsection, we consider the wireless communication is reliable and all data can be transmitted with high fidelity, i.e., the data quality is limited by the sensor precision rather than data communication. 

\begin{figure}[t]
\centering
    \includegraphics[width=0.48\textwidth]{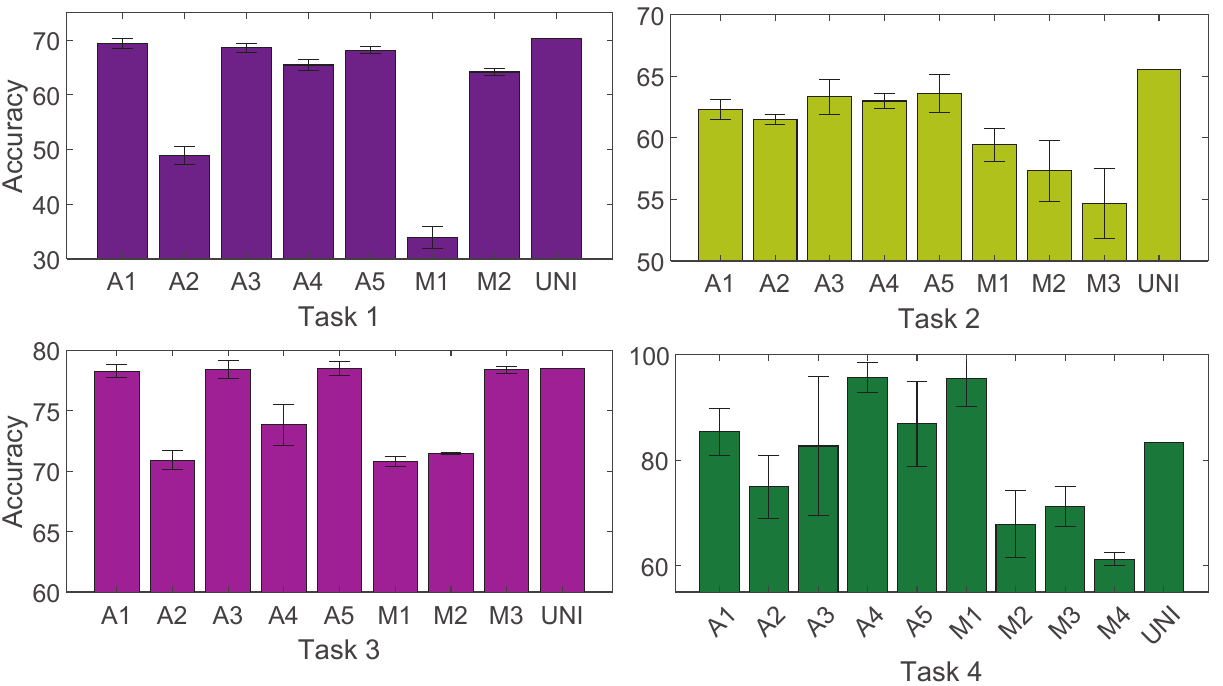}
    \vspace{-5pt}
    \caption{ Classification accuracy for 4 tasks using approaches A1 to A5. The single sensory modality performance is labeled as M1 to M4. UNI represents the HighMMT model for uni-task using all available sensory modalities in the associated task. }
    \label{fig:no_noise_acc}
    \vspace{-5pt}
\end{figure}

\begin{figure}[t]
\centering
    \includegraphics[width=0.35\textwidth]{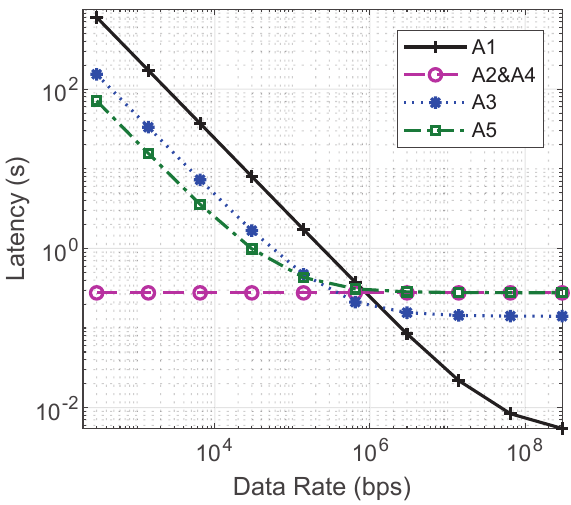}
    \vspace{-5pt}
    \caption{End-to-end latency using approaches A1 to A5 with different communication data rates for Task 1.}
    \label{fig:latency}
    \vspace{-5pt}
\end{figure}

\begin{figure}[t]
\centering
    \includegraphics[width=0.35\textwidth]{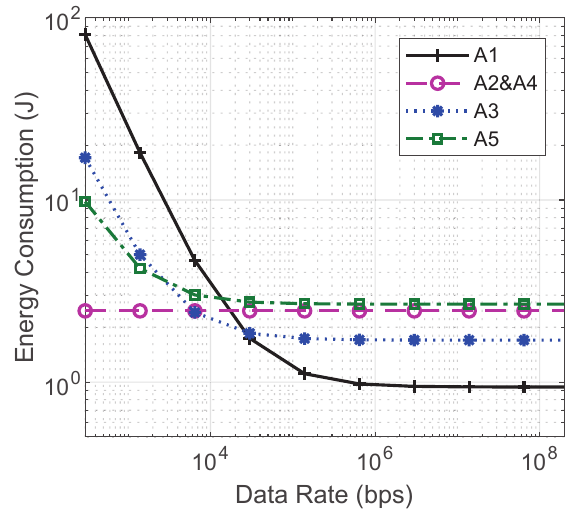}
    \vspace{-5pt}
    \caption{Energy consumption using approaches A1 to A5 with different communication data rates for Task 1.}
    \label{fig:energy}
    \vspace{-5pt}
\end{figure}

\subsubsection{Analysis of Sensory Modalities and Model Performance}

In Fig.~\ref{fig:no_noise_acc}, the classification accuracy of each task using approaches A1 to A5 are shown. For Conformal Prediction $\alpha$ is set as 0.1 and the percentile threshold for adaptive model selection is obtained using a percentage of 30\% for correctly classified calibration samples. {In order to better understand the performance, we also provide M1 to M4 and the UNI model. The single sensory modality performance for each task is labeled as M1 to M4. For example, Task 1 only has two sensory modalities and the accuracy is obtained using a unified single modality model (trained using 6 modalities with all 12 sensory data) for 4 tasks. UNI represents the performance of the HighMMT model trained for uni-task using all available sensory modalities in the associated task.}

The results reveal that each sensory modality has different task-related information. For example, M2 in Task 1 has much higher accuracy than M1. The uni-task (UNI) model for Task 1 performs slightly better than only using sensory modality 2. This shows that the sensory modalities are imbalanced in providing task-related information. Although deep learning models can extract underlying information and outperform single modality performance, e.g., Task 1 and Task 2, they can also perform similarly as the single modality model (e.g., Task 3). 

In A2, all sensory modalities are considered equally important. Due to the imbalanced task-related information, the low-accuracy sensory modalities will create negative impact on the final decision. As a result, A2 always performs the worst. A1 and A3 have similar performance except for Task 4. They are not exactly the same due to the limitation of GPU memory and we used a reduced test data size. Note that, A1 is the same as the HighMMT in \cite{liang2022high}. We notice that for all 4 tasks either A3 or A4 can achieve the best performance and A5's performance is in between A3's and A4's since it is a combination of them. Overall, A4 (MuSeCo) and A5 (MuSeCo-Adaptive) can achieve the highest average accuracy of 74.5\% and 74.3\%, which are similar to A1 (73.9\%) and A3 (73.3\%) and significantly better than A2 (64.1\%).

\subsubsection{Latency and Energy Consumption Evaluation}

To evaluate the latency and energy consumption, we consider the AIoT device is a typical microcomputer, such as Raspberry pi3, with energy efficiency $\Gamma_{iot}=0.813$ GFLOPS/W and computing capacity $c_{iot}=3.62$ GFLOPS. The edge server is a powerful machine with $\Gamma_{s}=2.13$ GFLOPS/W and computing capacity $c_s=428$ GFLOPS, e.g., Intel E5-2640 v3. We use fvcore library to count the number of FLOPS of the unimodal encoder and the multimodal cross attention module. They have the same structure and we consider the number of FLOPS are the same which is around 500 MFLOPS. The transmission power $P_t$ is 100 mW. {Since the transmission power is a constant, as the data rate increases, the approaches send high volume data demonstrates reduced overall energy consumption due to the shorter communication time.} The data rate using 64QAM modulation is shown as the x-axis in Fig.~\ref{fig:latency} and Fig.~\ref{fig:energy}. Since all the 4 tasks have similar latency and energy consumption performance, we only use Task 1 as an example. In the simulation, the average percentage ($p_h$) of samples using Conformal Prediction in A5 is 53.71\%.

As shown in Fig.~\ref{fig:latency} and Fig.~\ref{fig:energy}, A1 is the best solution if there is no limitation on data communication, i.e., the data rates can be high enough (larger than 1.5 Mbps). The latency and energy consumption of performing computations in the AIoT device is higher than that in the edge server. Therefore, it is faster and more energy-efficient to offload all computations to the edge server when data communication is fast. However, since the datasets have low-quality images and videos (e.g., 28 $\times$ 28), the required data rate is low. If the task requires high-resolution images and videos, the required data rates can be much higher than 1.5 Mbps. 

When the data rate is lower than 400 kbps, A2 and A4 have the lowest latency. In this scenario, the overall latency is constrained by communication latency. Since A2 only sends classification results and A4 sends Softmax scores and Conformal Prediction sets, the communication latency is low compared with sending raw data and latent data. Between 400 kbps and 1500 kbps, A3 has the lowest latency. In this data rate interval, the communication latency is smaller than sending raw data. Moreover, the offloading of multimodal cross attention computation to the edge server also reduces computation latency. The latency of A5 is between A3 and A4 when the performance is constrained by communication. As the available data rates increase, the latency caused by sending latent data and multimodal cross attention computation in the edge server are negligible and the latency of A5 converges to that of A2 and A4.  

Similarly, when the data rate is low, the energy consumption of communication is high and A2 and A4 have the lowest energy consumption, as shown in Fig.~\ref{fig:energy}. Similar to the latency, as data rates increase, A3 and A1 have the lowest energy consumption. This is mainly due to the high energy efficiency of the edge server and more computation workload is offloaded to the edge server. 

From the above discussions, we can see that A4 (MuSeCo) demonstrates high accuracy, low latency, and low energy consumption when the system is constrained by data communication. There is a transition interval in data rate where A3 performs better than any other approaches. When the system is not constrained by data communication, A1 can achieve the lowest latency and energy consumption. Note that, the exact data rates of the transition interval depend on the specific task and raw data size. Task-dependent communication and computation approaches can further improve the performance.

\subsection{Unreliable Communication}

Although 5G and Beyond wireless systems have significantly improved communication reliability and data rates, the AIoT networks still experience unreliable transmission occasionally due to interference, limited infrastructure support, and attacks. {In the following, we evaluate two scenarios. First, a single sensory modality is affected by unreliable communication and all other sensory modalities are reliably received by the edge server. Second, all sensory modalities are affected by unreliable communication. Here, we consider Additive White Gaussian Noise (AWGN) channel.}

\begin{figure}[t]
\centering
    \includegraphics[width=0.48\textwidth]{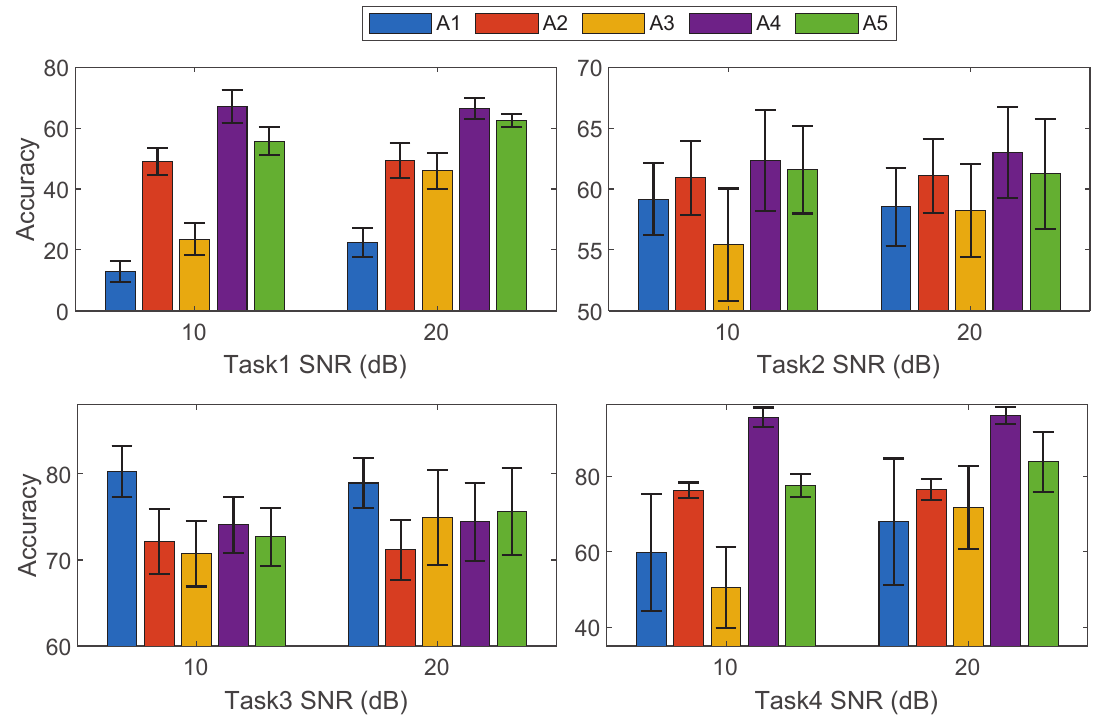}
    \vspace{-5pt}
    \caption{ Impact of imbalanced sensory modalities due to unreliable data communication. }
    \label{fig:imbalanced_noise}
    \vspace{-5pt}
\end{figure}

In all 4 tasks, the first sensory modality M1's SNR is set to 10 and 20 dB and all other sensory modalities' SNR are set to 30 dB. As we can see in Fig.~\ref{fig:imbalanced_noise}, a single modality with high noise power can affect the overall accuracy compared with the results in Fig.~\ref{fig:no_noise_acc}. First, A1's accuracy is significantly reduced since the raw sensory data is highly noisy which makes it challenging to classify the sample. Since the model is trained using clean data, the impact of noise can dramatically change the accuracy. For example, the accuracy of Task 1 is reduced from around 70\% to around 20\%. The accuracy of A1 for Task 3 is stable since M1 in Task 3 does not have strong task-related information compared with M3. Since A2 sends unimodal classification results and any error in data communication can cause significant errors, its accuracy is also affected by the noisy channel. But the impact is mitigated when there are many sensory modalities, e.g., Task 4 has 4 sensory modalities and A2's accuracy remains stable. In A3, latent data is sent to the edge server. Similar to A1, the noisy latent data of M1 creates negative impact on the multimodal cross attention and significantly reduces the classification accuracy. Although the accuracy of A4 and A5 are also affected by the noisy channel, the performance is among the best or close to the best in Task 1, 2 and 4. Also, due to the noise, the classification accuracy has a larger variance compared with the results in Fig.~\ref{fig:no_noise_acc}.

\begin{figure}[t]
\centering
    \includegraphics[width=0.48\textwidth]{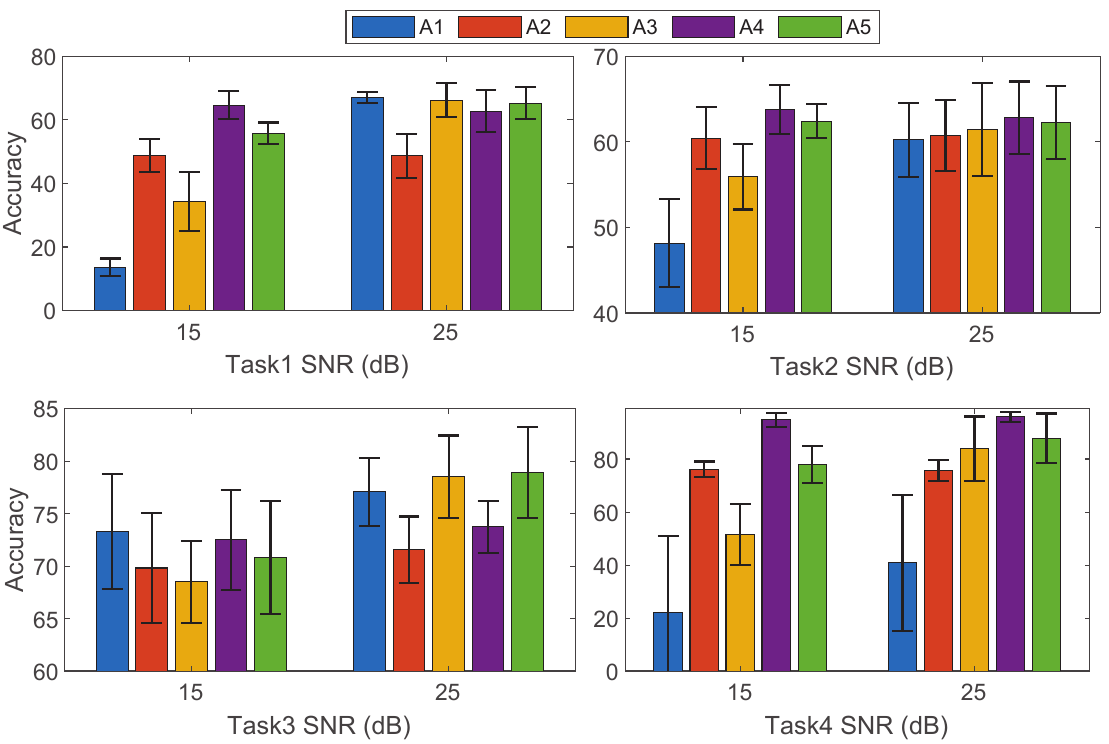}
    \vspace{-5pt}
    \caption{Impact of noisy sensory modalities due to unreliable data communication. }
    \label{fig:with_noise}
    \vspace{-5pt}
\end{figure}

In Fig.~\ref{fig:with_noise}, we consider two SNRs, i.e., 15 and 25 dB, for all sensory modalities. When SNR is 25 dB, we can consider the performance is similar to the reliable communication shown in Fig.~\ref{fig:no_noise_acc}. As we can see, A4 and A5 remain to be two reliable solutions without significant reduction in accuracy among the four tasks. The results for Task 3 is slightly different from other tasks. In particular, A1's accuracy decreases but not as significant as other tasks. A1 and A3 experience significant drop since the data is noisy. Although A2 performs better than A1 and A3 at low SNR, it cannot intelligently combine multimodal sensory data and its performance is worse than A4 and A5. 

In general, when the data communication is reliable A1, A3, A4, and A5 have similar performance in task accuracy, but A1, A3, and A4 demonstrate low latency and low energy consumption at different communication data rates. However, considering that A4's accuracy in reliable communication systems is slightly lower than that of A3 in 3 out of the 4 considered tasks (see Fig.~\ref{fig:no_noise_acc}), A5 can be a substitution of A4 to obtain higher accuracy with longer latency and higher energy consumption. With noisy communication channels, A4 and A5 demonstrate better performance compared with other solutions.

\section{Conclusion}

The Artificial Intelligence of Things (AIoT) enhances data fusion and communication efficiency, but the need for varied machine learning (ML) models to process different sensory modalities complicates the system and challenges large-scale deployment. In our study, we introduce MuSeCo and MuSeCo-Adaptive for AIoT, employing a unified Perceiver model and Conformal Prediction. This approach allows AIoT devices equipped with various sensors to use a single Perceiver model for data encoding. Conformal Prediction offers a lightweight solution for multimodal data fusion, which can select and integrate imbalanced multimodal data and reduce communication overhead. The model is trained across six sensory modalities for four distinct tasks. Evaluation results reveal that MuSeCo achieves high accuracy, minimal end-to-end latency, and low energy consumption in systems constrained by data communication. 

{The proposed frameworks open new avenues for research and development in AIoT systems. Future work could explore the integration of additional modalities, the extension of these frameworks to edge computing environments, and the application of MuSeCo in emerging domains such as autonomous vehicles and environmental monitoring. Furthermore, the principles underlying could inspire the development of similar unified frameworks for other complex, multimodal systems.}


\bibliographystyle{IEEEtran}
\bibliography{ref}







\end{document}